%% file: 0-Main.tex
\documentclass[acmsmall]{acmart}

\setcopyright{none}

\usepackage{parcolumns}
\usepackage[colorinlistoftodos,textwidth=15mm,textsize=tiny,obeyFinal]{todonotes}
\usepackage{subcaption}
\usepackage{mathtools} 
\usepackage{color}
\usepackage{textcomp}
\usepackage{etoolbox}
\usepackage{graphicx}
\usepackage{colortbl}
\usepackage{framed}
\usepackage{tabularx}
\usepackage[linesnumbered,ruled, vlined]{algorithm2e}
\usepackage{multirow,bigdelim}
\usepackage{enumitem}
\usepackage{balance}
\usepackage{url}
\usepackage[utf8]{inputenc}
\usepackage[labelfont=bf,tableposition=top]{caption}
\usepackage{booktabs}
\usepackage{threeparttable}
\usepackage{multirow}
\usepackage{adjustbox}
\usepackage{xcolor}
\usepackage{listings}
\usepackage{tcolorbox}
\usetikzlibrary{decorations,calligraphy}
\usepackage{fancyvrb}

\usepackage[english]{babel}
\usepackage{amsthm}

\theoremstyle{definition}
\newtheorem{definition}{Definition}[]

\AtBeginDocument{%
  }

\input{0-macro}

\begin{document} 

\title{Understanding and Detecting Annotation-Induced Faults of Static Analyzers}

\author{Huaien Zhang}
\orcid{0000-0001-6498-5062}
\affiliation{
  \institution{Hong Kong Polytechnic University}
  \city{Hong Kong}
  \country{China}
}
\affiliation{
  \institution{Southern University of Science and Technology}
  \city{Shenzhen}
  \country{China}
}
\email{cshezhang@comp.polyu.edu.hk}

\author{Yu Pei}
\orcid{0000-0001-6065-6958}
\affiliation{
  \institution{Hong Kong Polytechnic University}
  \city{Hong Kong}
  \country{China}
}
\email{yupei@polyu.edu.hk}

\author{Shuyun Liang}
\orcid{0009-0007-5746-8519}
\affiliation{
  \institution{Southern University of Science and Technology}
  \city{Shenzhen}
  \country{China}
}
\email{liangsy2022@mail.sustech.edu.cn}

\author{Shin Hwei Tan}
\orcid{0000-0001-8633-3372}
\affiliation{
  \institution{Concordia University}
  \city{Montreal}
  \country{Canada}
}
\email{shinhwei.tan@concordia.ca}

\begin{abstract}
Static analyzers can reason about the properties and behaviors of programs and detect various issues without executing them. Hence, they should extract the necessary information to understand the analyzed program well. Annotation has been a widely used feature for different purposes in Java since the introduction of Java 5. Annotations can change program structures and convey semantics information without awareness of static analyzers, consequently leading to imprecise analysis results. This paper presents the first comprehensive study of annotation-induced faults (AIF) by analyzing \issuecnt{} issues in \numtool{} open-source and popular static analyzers (i.e., \pmd{}, \sbugs{}, \cstyle{}, \infer{}, \sonar{}, and \soot{}). We analyzed the issues' root causes, symptoms, and fix strategies and derived \findingcnt{} findings and some practical guidelines for detecting and repairing annotation-induced faults. Moreover, we developed an automated testing framework called \sys{} based on three metamorphic relations originating from the findings. \sys{} generated new tests based on the official test suites of static analyzers and unveiled \faultcnt{} new faults, \fixedfaultcnt{} of which have been fixed. The results confirm the value of our study and its findings.
\end{abstract}

\maketitle

\input{1-Intro}
\input{2-Background}
\input{3-Methodology}
\input{4-AIFProneAnnotations}
\input{5-RootCause}
\input{6-Symptom}
\input{7-FixStrategy}
\input{8-Approach}
\input{9-Implication}
\input{10-Threat}
\input{11-RelatedWork}
\input{12-Conclusion}

\bibliographystyle{ACM-Reference-Format}
\bibliography{reference}

\end{document}

%% file: 0-macro.tex
\definecolor{main-color}{rgb}{0.6627, 0.7176, 0.7764}
\definecolor{string-color}{rgb}{0.3333, 0.5254, 0.345}
\definecolor{key-color}{rgb}{0.8, 0.47, 0.196}
\lstdefinestyle{mystyle}
{
    language = SQL,
    basicstyle = {\ttfamily \color{main-color}},
    stringstyle = {\color{string-color}},
    keywordstyle = {\color{key-color}},
    keywordstyle = [2]{\color{lime}},
    keywordstyle = [3]{\color{yellow}},
    keywordstyle = [4]{\color{teal}},
    morekeywords = [3]{<<, >>},
    morekeywords = [4]{++},
    basicstyle=\ttfamily\footnotesize,
    commentstyle=\color{blue}\ttfamily,
    morecomment=[f][\lstbg{red!20}]-,
    morecomment=[f][\lstbg{green!20}]+,
    morecomment=[f][\textit]{@@},
    numbers=left,
    texcl=false
}

\definecolor{pgrey}{rgb}{0.46,0.45,0.48}

\lstdefinestyle{javastyle}
{
    language = Java,
    basicstyle = {\ttfamily \color{main-color}},
    stringstyle = {\color{string-color}},
    keywordstyle = {\color{key-color}},
    keywordstyle = [2]{\color{lime}},
    keywordstyle = [3]{\color{yellow}},
    keywordstyle = [4]{\color{teal}},
    morekeywords = [3]{<<, >>},
    morekeywords = [4]{++},
    morekeywords = {List, Object},
    basicstyle=\ttfamily\scriptsize,
    commentstyle=\color{blue}\ttfamily,
    morecomment=[f][\lstbg{red!20}]-,
    morecomment=[f][\lstbg{green!20}]+,
    morecomment=[f][\textit]{@@},
    numbers=left,
    xleftmargin=4mm,
    numbersep=7pt,
    numberstyle=\scriptsize,
    frame=none,
    texcl=false,
    moredelim=[il][\textcolor{pgrey}]{\$\$},
    moredelim=[is][\textcolor{pgrey}]{\%\%}{\%\%}
}

\definecolor{grey}{rgb}{0.7,0.7,0.7}

\newcommand{\lstbg}[3][0pt]{{\fboxsep#1\colorbox{#2}{\strut #3}}}
\lstdefinelanguage{diff}{
  basicstyle=\ttfamily\footnotesize,,
  morecomment=[f][\lstbg{red!20}]-,
  morecomment=[f][\lstbg{green!20}]+,
  morecomment=[f][\textit]{@@},
  texcl=false
}

\def\sys{{\sc AnnaTester}\xspace}

\newcommand{\allissuecnt}{308}
\newcommand{\issuecnt}{246}
\newcommand{\faultcnt}{43}
\newcommand{\fixedfaultcnt}{20}
\newcommand{\fpcnt}{eight}
\newcommand{\rootcausecnt}{six}
\newcommand{\symptomcnt}{four}
\newcommand{\fixstrategycnt}{seven}
\newcommand{\findingcnt}{ten}

\newcommand{\numtool}{six}
\newcommand{\sonar}{SonarQube\xspace}
\newcommand{\pmd}{PMD\xspace}
\newcommand{\sbugs}{SpotBugs\xspace}
\newcommand{\cstyle}{CheckStyle\xspace}
\newcommand{\infer}{Infer\xspace}
\newcommand{\soot}{Soot\xspace}

%% file: 1-Intro.tex
\section{Introduction}
\label{sec:introduction}

Static analyzers are widely used to reason about programs and detect various issues without dynamically executing them.
Since they do not need to actually run the programs under consideration, static analyzers are highly applicable in a wide range of situations and have been utilized to reason about the properties and behaviors of programs in various development stages~\cite{tse2022usestaticanalysis}.
Meanwhile, since the analyses are based on the information they extract from program code, the static analyzers must have a good understanding of the syntax, semantics, and interplay of various constructs in the programs for the analysis results to faithfully reflect the real issues.
Since it is highly challenging to correctly handle all the program constructs in a static analyzer,
various techniques and tools have been developed in the past few years to detect bugs induced by incorrect handling of program constructs with predefined semantics in static analyzers~\cite{2022icpcfinder,cuoq2012testing,taneja2020testing,klinger2019differentially}.
However, many modern programming languages (e.g., Java, Python, and C\#) provide native support for annotations with programmer-defined semantics, 
which presents extra challenges to the reliability of static analyzers.

In computer programming, annotations are a form of syntactic metadata that associates additional information to various program elements.
Annotations have been utilized both as a more structured way to comment on code elements and as a mechanism to support meta-programming~\cite{CzarneckiEisenecker00}:
In the former case, they do not affect the semantics of programs;
In the latter case, they usually trigger special processing of the annotated elements during program compilation and/or execution, effectively extending the capabilities of the programming languages.
For instance, popular frameworks like Spring, JUnit, and Lombok rely heavily on their homebrewed annotations to simplify reusing the frameworks.
Nowadays, annotations have been widely utilized in practical software development.
According to a previous study conducted on a large number of open-source projects hosted on GitHub, the median number of annotations per project is up to 1,707~\cite{tse2019annotation}. 

In general, the presence of annotations poses two challenges to the reliability of static analyzers.
First, annotations in programs lead to extra tokens that static analyzers need to parse, while an unprepared static analyzer may overlook or mishandle the tokens, leading to incorrect analysis results or even premature termination of the tool.
For example, given the simple Java class in Figure~\ref{fig:motivatingexample} as input, the \pmd{} static analyzer will crash because the tool is not expecting the array access operators in line 5 to be annotated.
Second, annotations in programs may introduce changes to the structure or behavior of the programs at compile or execution time.
Since the detailed changes are defined by annotation processors, which are programs themselves, 
it is impractical to fully understand the impact of all annotations without running those processors,
and correspondingly, it is inevitable that static analyzers produce incorrect results if the annotations interfere with the programs' properties and behaviors being analyzed.

\begin{figure}[!htbp]
\begin{lstlisting}[style=javastyle]
@Retention(RetentionPolicy.CLASS)
@Target({TYPE_USE})
@interface Anno {}
public class Main {
  public <T> T[][] check(T @Anno[] @Anno [] arr) { // Trigger a crash
    if (arr == null) { 
      throw new NullPointerException();
    }  ...
\end{lstlisting}
\caption{An annotated Java program that will cause \pmd{} to crash.}
\label{fig:motivatingexample}
\end{figure}

To gain a better understanding of the extent to which annotations in programs affect the reliability of static analyzers,
we conducted the first large-scale empirical study on annotation-induced faults (AIFs) of static analyzers.
Although prior work has studied the usage and evolution of program annotations~\cite{tse2019annotation,rocha2011annotations, icsme2022mining, saner2022deepanna}, 
the maintenance of testing-related annotations~\cite{icse2021annotationmaintenance}, and the design of annotations for special purposes~\cite{clss2014java, jtres2010static}, there is little to no study on AIFs in static analyzers. 
This work aims to fill this gap.
Particularly, our study aims to answer the research questions below:
\begin{itemize}[leftmargin=*]

    \item {\textbf{RQ1:}} What \emph{kinds of annotations} are more likely to induce faults? 
    In RQ1, we study annotations that require attention when designing static analyzers. 

    \item {\textbf{RQ2:}} What are the \emph{root causes} for annotation-induced faults in static analyzers? In RQ2, we study the reasons behind annotation-induced faults to prevent them from reoccurring in the future. 
    
    \item {\textbf{RQ3:}} What are the \emph{symptoms} of those annotation-induced faults? In RQ3, we investigate the consequences of annotation-induced faults, which helps us assess the significance of the faults.
    
    \item {\textbf{RQ4:}} What are the \emph{fix strategies} that developers employ when fixing the annotation-induced faults?
    In RQ4, we strive to establish a good understanding of viable ways to fix annotation-induced faults, which is essential for reducing debugging efforts. 
    
\end{itemize}

To address the research questions, we manually analyzed \issuecnt{} annotation-induced issues and their corresponding patches from six popular open-source static analyzers, namely \pmd{}, \sbugs{}, \infer{}, \cstyle{}, \sonar{}, and \soot{}.
As a result, we uncovered \rootcausecnt{} main reasons for the annotation-induced faults, identified \symptomcnt{} symptoms of those faults, and unveiled \fixstrategycnt{} strategies developers adopted to fix the faults. 
We made \findingcnt{} major findings from the analysis results and discussed their implications for avoiding similar faults in the future. 
Based on our findings, we developed a framework named \sys{} to automatically detect three types of annotation-induced faults in static analyzers via metamorphic testing.
On the six aforementioned static analyzers, \sys{} successfully detected \faultcnt{} faults that were revealed for the first time.
We have reported the faults to the corresponding tool developers, and \fixedfaultcnt{} of them have been fixed at the time of writing, 
which clearly demonstrates the value of the framework, our study, and the findings.

In summary, this paper makes the following contributions:
\begin{itemize}[leftmargin=*]
    \item To the best of our knowledge, we conducted the first empirical study on annotation-induced faults in static analyzers based on \issuecnt{} issues from six popular open-source analyzers. We analyzed their root causes, symptoms, fix strategies, and types of annotations, deriving \findingcnt{} findings.
    \item Based on the findings from our study, we propose \sys{}, a new automated testing framework that uses metamorphic testing with our customized annotated program generator to detect three types of annotation-induced faults in static analyzers.
    \item We evaluated \sys{} on six static analyzers, and it was able to reveal \faultcnt{} new bugs in these static analyzers, \fixedfaultcnt{} of which have been confirmed and fixed. The experimental data and source code of \sys are available at: https://annaresearch.github.io/.
\end{itemize}

%% file: 2-Background.tex
\section{Background}
\label{sec:background}

\subsection{Static Analyzer}
\label{sec:bg_analyzer}
\begin{figure}[!tbp]
\centering
\includegraphics[width=\linewidth]{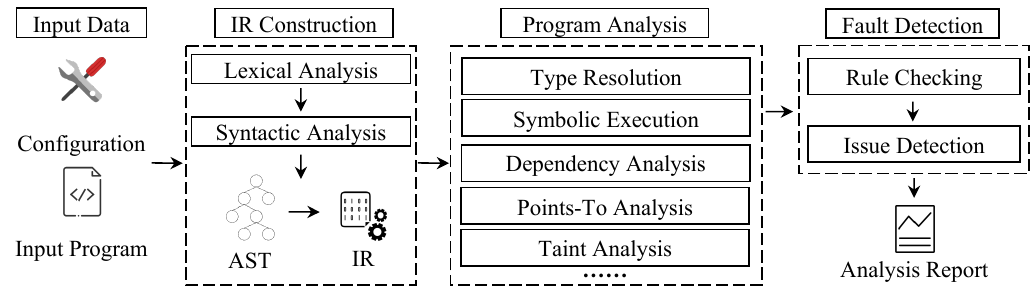}
\caption{The general workflow of a static analyzer}
\label{fig:saworkflow}
\end{figure}

Static analyzers are widely used to detect common issues without running programs.  Figure~\ref{fig:saworkflow} shows the general workflow of static analyzers based on previous work~\cite{sonarinaction, pmdworkflow, inferworkflow}. 
Using the program source code and a configuration as the input, 
a static analyzer first parses the program code and constructs an intermediate representation (IR).
Then, it applies different program analysis techniques like dependency analysis, symbolic analysis, etc., to extract relevant semantic information
from the program.
Finally, it employs rule checkers to detect issues based on the extracted information and reports the detected issues as the analysis result.

\subsection{Java Annotation}
\label{sec:bg_annotation}
Java annotations offer a structured way to attach helpful information to program elements like classes, methods, variables, and types.
An annotation essentially contains a (possibly empty) list of property-value pairs, with the properties specified in the annotation's definition and the values associated with the properties when each annotation declaration is used to annotate a program element.
Java annotations have no effect on the program semantics in and of themselves, but a program could implement annotation processors to adjust the program's code or even behaviors based on the presence of specific annotations, effectively extending the capability of the Java programming language.
For instance, an annotation \textit{@java.lang.Override} will trigger a compile-time check on the existence of the method it annotates in the current class's superclass, and an error will be issued if the check fails.

Java programs may use \emph{meta-annotations} (i.e., annotations applicable to other annotations) to restrict the application and effect of other annotations.
For example, meta-annotation \textit{@Target} is used to specify the types of elements (e.g., Class, Method, and Field) to which an annotation can be applied, while meta-annotation \textit{@Retention} is used to stipulate how long an annotation should be retained during a program's lifecycle: 
Source level retention means an annotation will not be retained after source code is compiled into bytecode;
Class retention means an annotation will be retained in bytecode but discarded when the bytecode is loaded into a JVM; 
Runtime retention means that an annotation is always retained and can be retrieved at runtime.

%% file: 3-Methodology.tex
\section{Empirical Study on Annotation-induced Faults}
\label{sec:methodology}

\subsection{Target Static Analyzers}
\label{sec:toolselection}

We select target analyzers based on three criteria: 
(1) it must be open-source and use a public issue tracking system (GitHub or Jira) to record all its issues that have been reported and resolved so that we can identify and analyze its AIFs and corresponding fixes; 
(2) it should be popular and widely used so that its issues are representative of the real problems faced by users of analyzers.
Particularly, we focus on analyzers with at least 2,000 stars on GitHub in this study; 
(3) it should support analysis of Java programs.
Based on these criteria, we select \numtool{} static analyzers:
(1) \pmd{}~\cite{pmdlink} is a cross-language static analyzer that detects common code smells, e.g., unused variables; 
(2) \sbugs{}~\cite{spotbugslink} is a fork of the now deprecated analyzer FindBugs that detects common bugs in Java programs via a set of code patterns; 
(3) \cstyle{}~\cite{checkstylelink} checks the conformity of Java code to a set of coding rules; 
(4) \infer{}~\cite{inferlink} is an analyzer designed by Meta to detect bugs for Java, C, C++, and Objective-C programs; 
(5) \sonar{}~\cite{sonarqubelink} is a continuous code inspection platform that detects bugs and code smells for various programming and markup languages;
(6) \soot{}~\cite{sootlink} is a static analysis framework that can analyze, instrument, and optimize Java and Android applications.

\subsection{Data Collection}
\label{sec:datacollection}

Among the \numtool{} target analyzers, \sonar{} uses Jira for tracking issues, while other tools use GitHub.
Since the issue tracking systems of these static analyzers contain around 14,000 issues, we refrained from manually inspecting all the issues to select only issues that are related to annotation-induced faults.
Instead, we first used the keyword ``annotation'' to search for closed issues that are likely annotation-induced.
We focus only on closed issues because how an issue was resolved sheds light on its root cause and fixing strategy, and we consider a closed issue relevant to annotations if and only if the keyword ``annotation'' appears in the issue's title or description.
The search returned \allissuecnt{} issues in total. 
Then, we manually checked these issues and excluded issues that were not associated with any fixing commits or not related to annotations.
Subsequently, we have \issuecnt{} faults to be analyzed in our study.
Table~\ref{tab:issuedistribution} lists the total number of issues for each analyzer in its issue tracking system (\#Issue$_{t}$), the number of likely annotation-induced issues returned by the keyword-based search (\#Issue$_{s}$), and the number of annotation-induced faults confirmed by the manual check and to be analyzed (\#Issue$_{a}$).
In the rest of this paper, we refer to faults using their IDs in the form TOOL-\#\#\#, where TOOL denotes the name of a static analyzer, while \#\#\# denotes the corresponding issue ID on GitHub or JIRA. As all issues are confirmed by the tool developers, we did not manually reproduce the collected issues.

\input{tab-IssueDistribution}

\subsection{Data Labeling and Reliability Analysis}

In this study, we identified the annotation-induced issues and analyzed them from three different aspects (i.e., the root cause, the symptom it exhibits, and the fix strategy).
The entire study took us around six months to complete.
To categorize (or label) the issues from each aspect, we followed previous work \cite{shen2021dlbugstudy, zhang2020dljob} to adapt existing taxonomies~\cite{sun2017mlbug, zhang2018tfbug, chen2019cloudexception, shen2021dlbugstudy, zhang2020dljob} to our task via an open-coding scheme. 
Specifically, one author first looked through all the issue reports and pull requests of those issues to determine the issue labels in these three aspects, including adding domain-specific categories and eliminating unnecessary categories. Then, two authors independently labeled the collected issues using the previously defined categories. 
We use Cohen's Kappa coefficient~\cite{cohen} to assess the agreement between these two authors. First, the two authors labeled 5\% of the issues, and Cohen's Kappa coefficient was nearly 0.69. Then, we had a training discussion and labeled 10\% of the issues (including the previous 5\%). 
At this stage, Cohen's Kappa coefficient reached 0.93. 
After an in-depth discussion on the issues with different labels, the two authors labeled the remaining issues in nine iterations, each covering ten more percent of the issues, and Cohen's Kappa coefficient remained greater than 0.9 in the process. In each iteration, the two authors discussed with the third author if they had any disagreement.
Finally, all the issues were labeled consistently.

%% file: tab-IssueDistribution.tex
\begin{table}[!htbp]
\caption{The issue distribution among six static analyzers}
\label{tab:issuedistribution}
\centering
\small
\begin{tabular*}{\linewidth}{@{\extracolsep{\fill} }lrrr}
\toprule
\textbf{Static Analyzer} & \#Issue$_t$ & \#Issue$_s$ & \#Issue$_a$ \\
\midrule
\sonar & 4370 & 138 & 128 \\
\cstyle & 4768 & 60 & 52 \\
\pmd & 2161 & 53 & 43 \\
\sbugs & 1043 & 10 & 7 \\
\infer & 1304 & 8 & 6 \\
\soot{} & 1147 & 39 & 10 \\
\midrule
\textbf{Total} & 14793 & \allissuecnt{} & \issuecnt{} \\
\bottomrule
\end{tabular*}
\end{table}

%% file: 4-AIFProneAnnotations.tex
\subsection{RQ1: AIF Prone Annotations}

\begin{figure}[!htbp]
\centering
\includegraphics[width=0.9\linewidth]{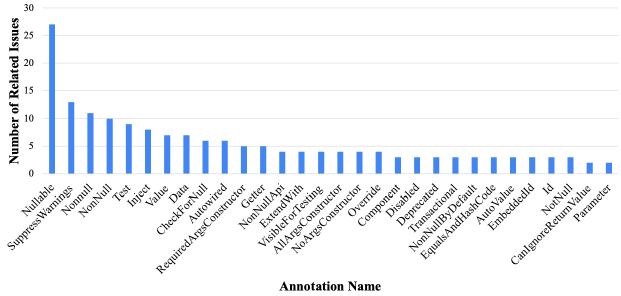}
\caption{The number of issues induced by each annotation from the top 30 most AIF-prone annotations.}
\label{fig:aifannotations}
\end{figure}

We collected annotations that trigger AIFs in the studied issues (we call them \emph{AIF prone annotations}) and sorted them in descending order of occurrence. 
Figure~\ref{fig:aifannotations} shows the top 30 most frequently occurred annotations in the study.
The x-axis shows the names of the annotations, and the y-axis presents the number of issues caused by each annotation. 

Overall, we observe that annotations that are used to specify the nullability of program elements (i.e., \textit{@Nullable},\textit{@Nonnull}, \textit{@NonNull}, \textit{@CheckForNull}, \textit{@NonNullApi}, and \textit{@NotNull}) are generally AIF prone annotations.
These nullability-related annotations triggered the most issues as static analyzers usually have many rules for checking the nullability of various program elements, 
and the implied semantics of these annotations may affect the analysis results.
Meanwhile, the annotation \textit{@SuppressWarnings} is often used to disregard specific warnings in static analyzers, e.g., \textit{@SuppressWarnings("WarningName")}. 
It caused the second most issues presumably because many static analyzers (e.g., \pmd{} and \sonar{}) support this annotation, and programmers often use this annotation for filtering out unwanted warnings.
Several test-related annotations (\textit{@Test},\textit{@ExtendWith} and \textit{@VisibleForTesting}) are AIF prone annotations due to their wide usages for marking tests.
Annotations \textit{@Inject} and \textit{@Autowired} support the automated injection of data dependence on annotated variables~\cite{aosd2009dataflow}.
Understandably, static analyzers may produce incorrect analysis results if they are unaware of the implicit data flow introduced by these annotations.
In Figure~\ref{fig:aifannotations}, around 23\% (i.e., 7/30) of the annotations (e.g., \textit{@Value}, \textit{@Data} and \textit{@Getter}) introduce changes to the original code, and failure to capture such changes may lead to bugs in static analyzers.

\begin{tcolorbox}[left=1pt,right=1pt,top=1pt,bottom=1pt]

\textbf{Finding 1:} 
Annotations that
(1) specify the nullability of program elements,
(2) are widely used (for marking unit tests or suppressing undesirable warnings), and
(3) alter the dependence or structure of the original code
have induced the largest number of faults in static analyzers.

\end{tcolorbox}

%% file: 5-RootCause.tex
\subsection{RQ2: Root Causes}
\label{sec:rootcause}

We uncovered a total of \rootcausecnt{} main reasons for the annotation-induced faults.
In this section, we use examples to explain the reasons in decreasing order of their total occurrences in our study.
Table~\ref{tab:rootcausedistribution} shows the number of faults caused by each reason for each static analyzer.

\input{tab-RootCauseDistribution}

\subsubsection{Incomplete Semantics (IS)}
\label{sec:semantics}
The most (38\%) common reason for AIFs  is that static analyzers usually only have incomplete knowledge about the annotations' semantics and, therefore, the semantics of the annotated programs.
Understandably, if an analyzer only has access to partial information it relies on, it is bound to produce inaccurate results.
Notably, no fault in \cstyle{} nor in \soot{} is related to this root cause because both tools largely ignore the semantics of the input program (\cstyle{} checks for coding styles, whereas \soot{} provides APIs for different analyses).

One fault induced by this root cause is SONAR-3804~\cite{sonar3804}, shown in Figure~\ref{fig:sq3804}.
\sonar has a rule stipulating that the keyword \texttt{volatile} should not be applied to non-primitive fields since when applied to a reference, the keyword makes sure that the reference itself, rather than the object it refers to, is never cached, which may cause obsolete object data to be cached and used by some program threads.
The stipulation, however, should be disregarded when the reference type's class is annotated with \textit{@Immutable} or \textit{@ThreadSafe}\footnote{Both annotations are defined in the package javax.annotation.concurrent.} since both annotations imply that the class's objects can be safely operated in multi-thread environments.
Being unaware of the semantics of the annotations, \sonar reported violations at lines 4 and 5 of the program in Figure~\ref{fig:sq3804}.

\begin{figure}[!htbp]
\begin{lstlisting}[style=javastyle]
@javax.annotation.concurrent.Immutable  class MyImmutable {}
@javax.annotation.concurrent.ThreadSafe  class MyThreadSafe {}
class Main {
    private volatile MyImmutable x; 
    private volatile MyThreadSafe y;
}
\end{lstlisting}
\caption{An incomplete semantics example in SONARQUBE-3804.}
\label{fig:sq3804}
\end{figure}

\begin{tcolorbox}[left=1pt,right=1pt,top=1pt,bottom=1pt]
\textbf{Finding 2:} 
Incomplete semantics is the most common root cause for faults in all studied analyzers, except for \cstyle{} and \soot{}. 
As annotations may introduce changes to the program properties and behaviors, failing to grasp the semantics encoded by the annotations will cause static analyzers to produce inaccurate results.
\end{tcolorbox}

\subsubsection{Improper AST Traversal (IAT)} 
\label{sec:iat}
After the developers of static analyzers have correctly constructed the abstract syntax tree (AST) for a program under syntactic analysis, 
they tend to misunderstand the impact of annotations on the ASTs and perform improper AST traversal.
In CHECKSTYLE-7522~\cite{cstyle7522}, the analyzer may encounter an ANNOTATION\_MEMBER\_VALUE\_PAIR node, i.e., a type of AST nodes used to represent the key-value pairs in annotation declarations like \textit{@Deprecated(removal=true)}, when it is not expecting one, which can cause a runtime crash.
In CHECKSTYLE-9941~\cite{cstyle9941}, an annotation for a method will push the nodes for the method's header comment one level down in the corresponding AST. Therefore, the static analyzer needs to access those nodes accordingly depending on the presence of annotations.
Failing to do that, \cstyle{} produced incorrect results when analyzing annotated methods.

\begin{tcolorbox}[left=1pt,right=1pt,top=1pt,bottom=1pt]
\textbf{Finding 3:} Developers of static analyzers tend to misunderstand the impact of annotations on ASTs, causing improper AST traversal to be the second most common root cause of  AIFs.
\end{tcolorbox}

\subsubsection{Unrecognized Equivalent Annotations (UEA)}
\label{sec:uea}
There are many annotations with equivalent semantics,
but developers of static analyzers often fail to recognize these annotations.
There are two types of equivalent annotations:
(1) Annotations from different libraries can have identical semantics and usage styles, but static analyzers often only recognize some of them, leading to inconsistent analysis reports. For example, the annotation \textit{@Nullable} means that the annotated element can hold a null value, and it has been supported in many popular third-party libraries (e.g., Google Android support, MongoDB, and Spring). Rules that check for null pointers need to analyze the program element annotated with \textit{@Nullable} to determine the nullability of the element. Moreover, with the dormition of JSR-305~\cite{jsr305dormant} (Java Specification Requests that aim to develop standard annotations for Java programs to assist software defect detection tools), several new libraries (e.g., Google JSR-305~\cite{googlejsr305}) have been proposed to implement annotations in JSR-305.
The annotations in these new libraries may cause UEA issues as they all comply with the same specification but have different names.
For example, while \sonar disallows variables of primitive data types to be declared as nullable in general, it reports a warning when \textit{@android.support.annotation.Nullable} is used to mark a boolean value as nullable but fails to do so when an equivalent annotation (i.e., \textit{@android.annotation.Nullable}) is used in the same way (lines 1 and 2 in Figure~\ref{fig:ueacase}).  
(2) As a library evolves, the fully qualified names of annotations defined in the library may also evolve.
For example, in SONARQUBE-3174~\cite{sonar3174}, the fully qualified package name of annotation \textit{@Generated} changed from \textit{javax.annotation} to \textit{javax.annotation.processing} but \sonar's developers were unaware of the change and failed to handle the renamed annotation correctly, causing inaccurate analysis results.

\begin{figure}[htbp]
\begin{lstlisting}[style=javastyle]
@android.support.annotation.Nullable boolean fun2() {} // report a warning
@android.annotation.Nullable boolean fun1() {}         // no warnings, an FN
\end{lstlisting}
\caption{SONARQUBE-3536~\cite{sonar3536}: A false negative caused by UEA}
\label{fig:ueacase}
\end{figure}

\begin{tcolorbox}[left=1pt,right=1pt,top=1pt,bottom=1pt]
\textbf{Finding 4:} Most (88\%) of the UEA faults were found in \sonar{}. Equivalent annotations may come from (1) different libraries or (2) different versions of the same library.
\end{tcolorbox}

\subsubsection{Erroneous Type Operations (ETO)} 
\label{sec:eto}
Several annotation-induced faults were due to erroneous type-related operations (e.g., missing type resolution, incorrect type casting/type checking).
In SONARQUBE-3438~\cite{sonar3438}, the \sonar{} developers mistakenly believed that the values stored in annotations can only be literals and incorrectly cast the AST for an annotation from ExpressionTree to LiteralTree, leading to a runtime exception.
Figure~\ref{fig:rctyperesolve} shows that in SONARQUBE-3045~\cite{sonar3045}, 
developers forgot to resolve the type of the annotation (i.e., \textit{@MyAnnotation}) applied on the actual type parameter (i.e., \textit{MyClass}), leaving the type parameter being annotated with an unknown type.

\begin{figure}[htbp]
\begin{lstlisting}[style=javastyle]
%%@Target(ElementType.TYPE_USE)%% %%@interface%% MyAnnotation {}
List<@MyAnnotation MyClass> field; // Unknown annotation type 
\end{lstlisting}
\caption{SONARQUBE-3045~\cite{sonar3045}: An incorrect type resolution in \sonar{}}
\label{fig:rctyperesolve}
\end{figure}

\subsubsection{Incorrect AST Generation (IAG)} 
\label{sec:iag}
Static analyzers may construct incorrect ASTs at the end of the IR construction stage (Figure~\ref{fig:saworkflow}).
One main reason for such problems is the grammar implemented by the static analyzers becomes obsolete after incorporating new rules about the usage of annotations into the specifications of new Java versions.
For example, the obsolete grammar prevented \cstyle{} from correctly parsing the annotations applied to the compact constructors of record types in CHECKSTYLE-8734~\cite{cstyle8734}. It also causes \sonar{} and \cstyle{} to incorrectly handle type annotations introduced by JSR-308 \cite{jsr308} in SONARQUBE-1420~\cite{sonar1420} and CHECKSTYLE-3238~\cite{cstyle3238}, respectively.
Meanwhile, some static analyzers may make mistakes in constructing ASTs from the tokens returned by the lexers.
For example, in SONARQUBE-1167~\cite{sonar1167}, although \sonar{} correctly extracted the annotations placed on type parameter declarations,
it failed to store the information correctly in the corresponding AST, leading to an incorrect AST.

\begin{tcolorbox}[left=1pt,right=1pt,top=1pt,bottom=1pt]
\textbf{Finding 5:} 
Incorrect AST generation is a common root cause, and most (85.7\%) of the IAG faults were due to the obsolete grammar that static analyzers implement.
\end{tcolorbox}

\subsubsection{Misprocessing of Configuration File (MCF)}
\label{sec:mcf}
As shown in Figure~\ref{fig:saworkflow}, a static analyzer usually expects as the input both the program files to be analyzed and a configuration file that specifies rules to be enabled and/or disabled, locations of the auxiliary libraries, etc.
Several faults occurred because static analyzers failed to process the given configuration files correctly.
For example, \pmd reads from the configuration file a list of annotations to be ignored in its analysis.
In PMD-2454~\cite{pmd2454}, developers forgot to trim the leading and trailing whitespaces when extracting annotation names from the configuration files, causing the failure to match ``\textit{@PreDestroy\textvisiblespace}'' with ``\textit{@PreDestroy}''.

\subsubsection{Others}
\label{sec:others}
Two faults were highly specific to their corresponding tool implementations and cannot be attributed to any of the aforementioned reasons.
In SONARQUBE-3108~\cite{sonar3108}, \sonar crashes with an OutOfMemory exception when analyzing a method with 24 parameters, all annotated with \textit{@Nullable}.
The reason is that \sonar creates two symbolic starting states (``NULL'' and ``NOT\_NULL'') for each nullable parameter, and 
it needed to create $2^{24}$ symbolic states for parameters of the method, which exceeded the available memory in the JVM.
In CHECKSTYLE-2202~\cite{cstyle2202}, \textit{@SuppressWarnings} is utilized to suppress warnings specified by the annotation parameters, but developers ignore the parameters named in camel-case notation, leading to a false positive.

%% file: tab-RootCauseDistribution.tex
\begin{table}[!htbp]
\caption{The numbers of faults due to different root causes in each analyzer.}
\label{tab:rootcausedistribution}
\setlength{\tabcolsep}{3pt}
\centering
\footnotesize
\begin{threeparttable}
\begin{tabular*}{\linewidth}{@{\extracolsep{\fill} }lrrrrrrrr}
\toprule
\textbf{Static Analyzer} & IS & IAT & UEA & ETO & IAG & MCF & Others & \textbf{Total} \\
\midrule
\sonar{} & 66 & 23 & 21 & 11 & 4 & 2 & 1 & 128 \\
\cstyle{} & 0 & 39 & 1 & 0 & 10 & 1 & 1 & 52 \\
\pmd{} & 19 & 9 & 0 & 4 & 7 & 4 & 0 & 43 \\
\sbugs{} & 4 & 0 & 0 & 3 & 0 & 0 & 0 & 7 \\
\infer{} & 4 & 0 & 0 & 1 & 0 & 1 & 0 & 6 \\
\soot{} & 0 & 3 & 2 & 4 & 0 & 1 & 0 & 10 \\
\midrule
\textbf{Total} & 93 & 74 & 24 & 23 & 21 & 9 & 2 & \issuecnt{} \\
\bottomrule
\end{tabular*}
\begin{tablenotes}[flushleft]
\footnotesize
  \item[] \textbf{IS}: {Incomplete Semantics}; \textbf{IAT}: Improper AST Traversal; \textbf{UEA}: Unrecognized Equivalent Annotations; \textbf{ETO}: Erroneous Type Operation; \textbf{IAG}: Incorrect AST Generation; \textbf{MCF}: Misprocessing of Configuration File.
\end{tablenotes} 
\end{threeparttable}
\end{table}

%% file: 6-Symptom.tex
\subsection{RQ3: Symptoms}
\label{sec:symptom}
In this section, we describe the four symptoms and then relate them to the \rootcausecnt{} root causes to help users and developers assess the impacts of different root causes.

\subsubsection{Symptom Category}
All symptoms caused by annotation-induced faults are listed below:

\begin{description}[leftmargin=*]
    \item[False Positive:] Symptoms in this category involve analysis reports with undesirable warnings.
    
    \item[False Negative:] Symptoms in this category involve analysis reports that are missing warnings. 

    \item[Crash/Error:] Symptoms in this category involve premature terminations or compilation errors. 
    
    \item[Other Wrong Results: ] 
    While most bug reports of the studied faults contain descriptions of the symptoms caused w.r.t.\ the final results produced by analyzers, 
    some reports only referred to incorrect intermediate results generated during the analyses without explaining how those intermediate results affect the overall analysis outcome.
    We classify those faults into this category.
    For instance, the issue report of CHECKSTYLE-8734~\cite{cstyle8734} explains that \cstyle{} cannot parse the annotation on Java records and will construct only a partial AST for the program under analysis.
\end{description}

Table~\ref{tab:symptomdistribution} shows the distribution of the four categories of symptoms across the six static analyzers. 
We observe that most of the studied faults fall into the false positive (FP) category, probably because these faults were discovered during the actual use of the analyzers, and users were more sensitive to undesirable warnings in analysis reports. The prevalence of FPs is also in line with the findings of prior studies on static analyzers~\cite{johnson2013don,ayewah2007evaluating,how_many,fixit,christakis2016developers}.
Note that we identify one symptom for each fault in our study based on the bug report, 
which is reasonable as each bug report usually focuses on only one particular negative impact.
Although it may happen in practice that a run of an analyzer exhibits multiple symptoms from different categories,
we can reliably make the following finding.

\begin{tcolorbox}[left=1pt,right=1pt,top=1pt,bottom=1pt]
\textbf{Finding 6:} 
Annotation-induced faults may cause static analyzers to produce inaccurate analysis results, to crash at runtime, or to generate incorrect intermediate results.
\end{tcolorbox}

\input{tab-SymptomDistribution.tex}

\subsubsection{Relationship between Root Causes and Symptoms}
So far, we have summarized the root causes and symptoms of annotation-induced faults. Understanding their relationship can help us better comprehend the impact of various root causes on static analysis results.
Table~\ref{tab:rcsrelationship} shows the relationship between the root causes and symptoms. 
Although IS was the most common root cause, we observed that it never triggered runtime crashes, which were mostly caused by IAG. 
IS only led to inaccurate analysis results, especially at the program analysis stage, while crashes often occurred at IR construction stage.
We also observe that while FP results can be triggered by all root causes of AIFs, FN and CE results were never caused by incomplete semantics.

\begin{tcolorbox}[left=1pt,right=1pt,top=1pt,bottom=1pt]
\textbf{Finding 7:} 
All identified root causes in our study led to FPs.
Incomplete semantics was the most common root cause and typically led to incorrect analysis results (i.e., FP).
\end{tcolorbox}

\input{tab-RootCauseSymptom.tex}

%% file: tab-SymptomDistribution.tex
\begin{table}[!htbp]
\caption{The number of issues for the four categories of symptoms across the static analyzers.}
\label{tab:symptomdistribution}
\centering
\setlength{\tabcolsep}{3pt}
{
\begin{threeparttable}
\begin{tabular*}{\linewidth}{@{\extracolsep{\fill} } lrrrrr}
\toprule
\textbf{Static analyzer} & FP & CE & FN & OWR & \textbf{Overall} \\
\midrule
\sonar{} & 99 & 7 & 12 & 10 & 128 \\
\cstyle{} & 26 & 14 & 10 & 2 & 52 \\
\pmd{} & 28 & 8 & 7 & 0 & 43 \\
\sbugs{} & 4 & 0 & 3 & 0 & 7 \\
\infer{} & 5 & 0 & 1 & 0 & 6 \\
\soot{} & 0 & 6 & 0 & 4 & 10 \\
\midrule
\textbf{Overall} & 162 & 35 & 33 & 16 & \issuecnt{} \\
\bottomrule
\end{tabular*}
\begin{tablenotes}
\footnotesize
  \item[] \textbf{FP}: False Positive, \textbf{CE}: Crash/Error, \textbf{FN}: False Negative, and \textbf{OWR}: Other Wrong Results.
\end{tablenotes} 
\end{threeparttable}
}
\end{table}

%% file: tab-RootCauseSymptom.tex
\begin{table}[!htbp]
\caption{Relationship between the root causes and symptoms of annotation-induced faults.}
\label{tab:rcsrelationship}
\centering
\setlength{\tabcolsep}{1pt}
\small
\begin{threeparttable}
\begin{tabular*}{\linewidth}{@{\extracolsep{\fill} } lrrrrrrrr}
\toprule
\textbf{Symptom} & IS & IAT & UEA & ETO & IAG & MCF & Others & \textbf{Overall} \\
\midrule
False Positive & 93 & 40 & 11 & 9 & 4 & 4 & 1 & 162 \\
Crash/Error & 0 & 8 & 2 & 8 & 14 & 2 & 1 & 35 \\
False Negative & 0 & 19 & 7 & 4 & 0 & 3 & 0 & 33 \\
Wrong Intermediate Result & 0 & 7 & 4 & 2 & 3 & 0 & 0 & 16 \\
\bottomrule
\end{tabular*}
\begin{tablenotes}
\footnotesize
  \item[] \textbf{IS}: {Incomplete Semantics}; \textbf{IAT}: Improper AST Traversal; \textbf{UEA}: Unrecognized Equivalent Annotations; \textbf{ETO}: Erroneous Type Operation; \textbf{IAG}: Incorrect AST Generation; \textbf{MCF}: Misprocessing of Configuration File.
\end{tablenotes} 
\end{threeparttable}

\end{table}

%% file: 7-FixStrategy.tex
\subsection{RQ4: Fix Strategies}
\label{sec:fixstrategy}

We unveiled \fixstrategycnt{} common fix strategies for fixing annotation-induced faults.
In this section, we first introduce each fix strategy and then relate the fix strategies to the root causes of the faults.

\subsubsection{Fix Incorrect Use of Annotation Filter (FAF)}

As there can be many programmer-defined annotations with distinct semantics, a static analyzer often utilizes white and black lists to filter the annotations that it will or will not support.
Such a list can be hard-coded into the static analyzer or fed to the static analyzer as part of a configuration file.
For example, the \textit{ignoredAnnotations} property in \pmd's configuration file is used to specify the annotations to be neglected by specific rule checkers. 
In general, annotation filtering may suffer from two types of problems.
First, a list may miss some annotations or contain undesirable annotations. 
For instance, in SONARQUBE-1513~\cite{sonar1513}, a rule checker was used to identify subclasses that should override the \textit{equals} method, 
but it mistakenly ignored the annotation \textit{@EqualsAndHashCode} in Lombok which, when applied to a class, will cause boilerplate implementations of \textit{equals} and \textit{hashCode} methods to be inserted into the class. 
To fix this fault, the developers added the annotation to the white list, and \sonar{} will default to the class with this annotation having overridden \textit{equals} method.
Second, the utilization of the lists may be faulty.
e.g., in PMD-2876~\cite{pmd2876}, a \pmd{} user specified the list of ignored annotations in the configuration file to customize the Lombok annotations to be neglected by the tool, 
but the customization failed due to \pmd's incorrect handling of the list.

\begin{tcolorbox}[left=1pt,right=1pt,top=1pt,bottom=1pt]
\textbf{Finding 8:} 
Incorrect use of annotation filters was fixed by adjusting the annotation lists in 90.2\% cases and correcting the mishandling of annotation lists in the remaining 9.8\% cases.
\end{tcolorbox}

\subsubsection{Fix AST Node Retrieval (FAN)}
The ASTs of programs under analysis are essential information for static analyzers,
but the process of information extraction from ASTs may suffer from two types of problems.
First, the analyzers may misunderstand the structure of the ASTs, especially when the annotations used in the programs introduce changes to the ASTs.
For instance, in CHECKSTYLE-10945~\cite{cstyle10945}, the tool developers mistakenly neglected the \textit{ARRAY\_INIT\_ARRAY} nodes as part of the annotations in the ASTs.
To fix such problems, programmers need to adjust their traversal algorithms based on the actual structure of ASTs.
Second, the computation performed by an analyzer when traversing an AST may be faulty.
For example, \pmd employs a flag variable named \textit{hasLombok} to track in a depth-first AST traversal whether a class has an annotation from the Lombok library, and it will suppress all the \textit{SingularField} warnings on classes where the variable value is true. 
In PMD-1641~\cite{pmd1641}, the traversal algorithm forgot to restore the variable's value to false after returning from the visit to an inner class, causing an unwanted \textit{SingularField} warning.
The code snippet in Figure~\ref{fig:incorrectalgo} shows how the fault was fixed.

\begin{figure}[!htbp]
\begin{lstlisting}[style=javastyle]
+ boolean tmp = hasLombok;
  hasLombok = hasLombokAnnotation(node);
  Object result = super.visit(node, data);
+ hasLombok = tmp;
\end{lstlisting}
\caption{PMD-1641~\cite{pmd1641}: Fix incorrect traversal algorithm}
\label{fig:incorrectalgo}
\end{figure}

\subsubsection{Fix Incorrect Type Operation (FIT)}
This strategy involves fixing erroneous type-related operations (e.g., type resolution and type casting).
For example, in SONARQUBE-2205~\cite{sonar2205}, the developer mistakenly resolved the type of an annotation based on its simple name, and the fix involves replacing the simple name with the annotation's fully qualified name.
In PMD-1369~\cite{pmd1369}, a runtime crash occurred due to an incorrect cast of a reference from type \textit{ASTAnnotation} to type \textit{ASTClassOrInterfaceType}. 
To fix this, a type compatibility check was added to guard the type casting.

\subsubsection{Fix Grammar Issue (FGI)}
As shown in Figure~\ref{fig:saworkflow}, static analyzers rely on predefined grammar to perform lexical and syntactic analysis to generate intermediate representative. We classify grammar-related fix strategies into two subcategories: (1) Fix lookahead parameter. Lookahead is often used in the lexical analysis stage. It can match the specific tokens in the source code to be analyzed. (2) Fix grammar patterns. Static analyzers can define grammar patterns to recognize corresponding syntax structures. However, these patterns may ignore annotations directly, or new usages of annotation, e.g., in CHEKCSTYLE-3238~\cite{cstyle3238}, developers did not define grammar patterns to recognize annotations on variable-length parameters and failed to parse them consequently.

\subsubsection{Fix Value Check (FVC)}
This fix strategy involves adding checks that were missing or rectifying checks that were inappropriate.
For example, in CHECKSTYLE-4472~\cite{cstyle4472}, a missing null value check caused a runtime crash, and the fix was to add the missing check.

\subsubsection{Redesign Rule Checker Pattern (RRC)}
Static analyzers based on rule checkers (e.g., all evaluated tools except for Soot) use predefined patterns to detect bugs, but the patterns may be incorrect and need to be redesigned. 
For instance, Figure~\ref{fig:redesignrule} shows that in PMD-1782~\cite{pmd1782} the rule checker initially only checks if a class or interface has a package definition (ignoring annotation).
In line 2, PMD checks whether a package definition exists by counting the number of occurrences of the \textit{PackageDeclaration} node in the XPath (which represents AST as an XML-like DOM structure) but mistakenly omitted the annotation when writing the XPath.
To fix this, developers redesign the rule in the XPath to check if a package declaration exists in the initial lines of a compilation unit.

\begin{figure}[!htbp]
\begin{lstlisting}[style=javastyle]
- /ClassOrInterfaceDeclaration[count(preceding::PackageDeclaration)=0]
+ CompilationUnit[not(./PackageDeclaration)]/TypeDeclaration[1]
\end{lstlisting}
\caption{PMD-1782~\cite{pmd1782}: Redesign rule pattern to recognize package declaration}
\label{fig:redesignrule}
\end{figure}

\subsubsection{Fix Incorrect API Usage (FIA)}
This fix strategy involves repairing incorrect API usages, mainly by using the correct API to retrieve elements or parse the signature of an annotation. Figure~\ref{fig:sootapi} shows that in SOOT-123~\cite{soot123} when creating an \textit{AnnotationTag} object, \soot{} incorrectly invoked the API \textit{DexType.toSoot} (line 1) to prepare a type descriptor as the actual parameter for invoking the \textit{AnnotationTag} constructor. Figure~\ref{fig:incorrectapi} shows that in CHECKSTYLE-2202~\cite{cstyle2202}, users adopt the \textit{@SuppressWarnings} annotation to disable a warning, but \cstyle{} only recognizes rule names in lower case (line 1) and fails to detect equivalent rule names in camel case (line 2). To fix this, developers use $equalsIgnoreCase$ instead of $equals$ to recognize all the equivalent rule names.

\begin{figure}[!htbp]
\begin{lstlisting}[style=javastyle]
- AnnotationTag aTag = new AnnotationTag(DexType.toSoot(a.getType()).toString());
+ AnnotationTag aTag = new AnnotationTag(a.getType());
\end{lstlisting}
\caption{SOOT-123~\cite{soot123}: Incorrectly invoke \textit{toSoot} to construct an \textit{AnnotationTag}}
\label{fig:sootapi}
\end{figure}

\begin{figure}[!htbp]
\begin{lstlisting}[style=javastyle]
@SuppressWarnings("checkstyle:redundantmodifier") // No warnings
@SuppressWarnings("checkstyle:RedundantModifier") // Report a warning, but it is an FP
\end{lstlisting}
\caption{CHECKSTYLE-2202~\cite{cstyle2202}: Failing to recognize the camel case leads to an FP}
\label{fig:incorrectapi}
\end{figure}

\subsubsection{Others}
Three faults were not fixed by previously discussed fix strategies. In INFER-559~\cite{infer559}, \infer{} reported an FP because it only used method signature information to analyze the parameter properties for compiled Java programs, while no annotation is retained in the method signature information, causing the tool to miss out on all annotations on method parameters.

\subsubsection{Relationship between Root Cause and Fix Strategy}

Knowledge about the relationship between root causes and fix strategies is valuable for guiding the fix of annotation-induced faults. 
Table~\ref{tab:relationrcfix} shows the number of annotation-induced bugs caused by each root cause and fixed with each strategy.
As shown in the table, FAF is the most commonly adopted fix strategy.
Moreover, most faults fixed by strategy FAF were due to root causes IS or UEA, probably because it was too challenging for the static analyzers to correctly handle the semantics of the annotations involved in those faults. Therefore, the tool developers resorted to the filter-based solution as a workaround for the faults.

\input{tab-RootCauseFix}

\begin{tcolorbox}[left=1pt,right=1pt,top=1pt,bottom=1pt]
\textbf{Finding 9:} FAF was the most common fix strategy, especially for faults caused by IS and UEA.
\end{tcolorbox}

FIT is also a popular fix strategy and can fix most root causes except for IS and UEA (both are fixed by FAF mostly). Most FIT related issues are due to fixing type resolution (15) and type checking (11). Sec.~\ref{sec:rootcause} states that type resolution issues are caused by incorrect auxiliary library configuration and missing identifier resolution. To fix the former type resolution issue, developers need to load proper libraries and find the correct class file to resolve the annotation. For the latter, developers should consider all possible program elements that need resolutions (e.g., the fault in Fig.~\ref{fig:fixtyperesolution} occurs because it fails to resolve the annotation in fully qualified name \textit{org.foo.@MyAnnotation}, leading to an unused import FP at line 2).
FGI mainly appears in one root cause (IAG) because incorrect grammar leads to parsing failure.
To fix them, developers should check whether the next token from the lexical stream is a ``@'' symbol. Based on our study, developers often made mistakes when handling annotations on throw type, variable arguments, and generic type.

\begin{figure}[!htbp]
\begin{lstlisting}[style=javastyle]
package org.foo;
import org.foo.bar.MyAnnotation; // report an FP
class A {
  org.foo.@MyAnnotation B myB;
}
\end{lstlisting}
\caption{SONARQUBE-2083~\cite{sonar2083}: Fail to resolve annotation fully qualified name}
\label{fig:fixtyperesolution}
\end{figure}

\begin{tcolorbox}[left=1pt,right=1pt,top=1pt,bottom=1pt]
\textbf{Finding 10:} Among all fix strategies, FIT covers the greatest number of root causes. Fixing type cast and type resolution account for the majority of issues. 
\end{tcolorbox}

%% file: tab-RootCauseFix.tex
\begin{table}[!htbp]
\caption{Relationship between root cause and fix strategy}
\label{tab:relationrcfix}
\centering\setlength{\tabcolsep}{1pt}
\small
\begin{tabular*}{\linewidth}{@{\extracolsep{\fill} }lrrrrrrrrr}
\toprule
\textbf{Root Cause} & FAF & FAN & FIT & FGI & FVC & RRC & FIA & Others & Overall \\
\midrule
Incomplete Semantics (\textbf{IS}) & 83  & 0 & 0 & 0 & 3 & 4 & 0 & 3  & 93 \\
Improper AST Traversal (\textbf{IAT}) & 7 & 36  & 13  & 0 & 13  & 5 & 0 & 0 & 74 \\
Unrecognized Equivalent Annotations (\textbf{UEA}) & 21  & 1 & 0 & 1 & 1 & 0 & 0 & 0 & 24 \\
Erroneous Type Operations (\textbf{ETO}) & 1 & 1 & 16  & 0 & 3 & 0 & 2 & 0  & 23  \\
Incorrect AST Generation (\textbf{IAG}) & 0 & 0 & 1 & 20  & 0 & 0 & 0 & 0  & 21  \\
Misprocessing of Configuration File (\textbf{MCF}) & 2 & 0 & 3 & 1 & 0 & 1 & 1 & 1  & 9 \\
Others & 0 & 0 & 0 & 1 & 0 & 0 & 1 & 0 & 2 \\
\hline
\textbf{Overall} & 114 & 38 & 33 & 23 & 20 & 10 & 4 & 4 & \issuecnt{} \\
\bottomrule
\end{tabular*}
\end{table}

%% file: 8-Approach.tex
\section{Implementation of \sys{} framework}
\label{subsec:cg}

\begin{figure}[!htbp]
\centering
\includegraphics{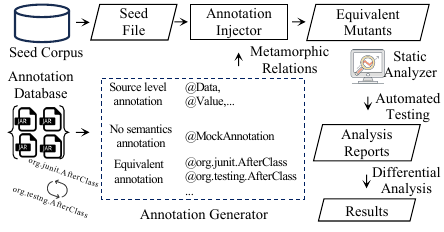}
\caption{Overall Workflow of \sys{}}
\label{fig:work}
\end{figure}

We propose \sys{}, a framework for automatically detecting annotation-induced faults via metamorphic testing, and it includes three checkers motivated by our study findings.
Figure~\ref{fig:work} shows the overall workflow of \sys{}. 
Given a set of input programs obtained from the test suite of a static analyzer under test as the input, 
\sys{} detects annotation-induced faults in the analyzer in two steps: 
(1) generates annotated programs by injecting annotations into the input programs;
(2) checks whether the analysis reports produced by the analyzer on the input programs, both with and without annotations, satisfy the corresponding metamorphic relations.
We adopt Eclipse JDT library to parse source seed files and inject annotations.

\subsection{Checkers and metamorphic relations}
\sys essentially relies on three checkers to detect AIFs, each of which is based on a metamorphic relation concerning analysis reports on programs with and without annotations.
When the metamorphic relation of a checker is violated, an AIF is detected, 
and the checker reports violations together with the input programs, with and without annotations, to users for further analysis.
All metamorphic relations are based on the analysis equivalence relation between programs as below:
\begin{definition}[Analysis Equivalence]
\label{def:cn}
Two programs $P$ and $P'$ are analysis equivalent w.r.t.\ a static analyzer $S$, denoted as $P\equiv_S P'$, if and only if (1) $S$ reports the same issues on $P$ and $P'$ and (2) $S$ terminates in the same state, i.e., successful or with errors when applied on $P$ and $P'$.
We use $P\equiv P'$ to denote that $P$ and $P'$ are analysis equivalent w.r.t.\ any static analyzer.
\end{definition}
In the rest of this subsection, we will use the following notations.
Let $P$ be a Java program, $a$ be an annotation, 
$\mathcal{P}(P)$ denotes the resultant program produced by processing the annotations in $P$;
$\mathcal{I}(P, a)$ denotes the set of all programs produced by applying $a$ to appropriate elements in $P$. 

\subsubsection{Incomplete Semantics Checker ({ISC})}
Incomplete semantics (IS) was the most common root cause for AIFs. 
In our study, one fault in \pmd due to IS was evidenced by contrasting the analysis reports produced on two programs that were supposed to be analysis equivalent since the second program is derived by processing all the annotations in the first one~\cite{diffmotivation}.
Motivated by this example and Finding 2, we design a metamorphic relation requiring that a program $P$ should be analysis equivalent to the resultant program produced by processing the annotations in $P$.

\begin{definition}[MR1]
\label{def:mr1}
Given a program $P$, $P$ and $\mathcal{P}(P)$ should be analysis equivalent, i.e., $P\equiv \mathcal{P}(P)$.
\end{definition}

As stated in Section~\ref{sec:bg_annotation}, source-level annotations are not retained in the compiled code (i.e., the semantics of those annotations must be fully processed and incorporated into the program code during compilation).
Hence, the differences between static analysis results of the programs before and after their source-level annotations have been processed indicate potential annotation-induced faults due to IS in the analyzers.
In view of that, the incomplete semantics checker focuses on detecting IS faults caused by source-level annotations.
Since IS never led to faults in \cstyle{} and \soot{} (Finding 2), we do not apply this checker to detect faults in these two analyzers.

\subsubsection{Annotation Syntax Checker ({ASC})}
Findings 3 and 5 indicate that incorrect AST generation and traversal may cause static analyzers to produce inaccurate analysis results or even crashes.
As such negative influences are independent of the semantics of the involved annotations, 
we implement an annotation syntax checker based on the following metamorphic relation on \emph{dummy annotations} (i.e., annotations that mark program elements but have no impact on programs' semantics).
Notably, metamorphic relation MR2 states that adding dummy annotations to a program should not affect the static analysis detection results produced on the program.

\begin{definition}[MR2]
\label{def:mr2}
Given a program $P$ and a dummy annotation $d$, $\forall p\in \mathcal{I}(P, d): P\equiv p$.
\end{definition}

\subsubsection{Equivalent Annotation Checker ({EAC})}
To identify inconsistent behaviors across equivalent annotations, we design the following metamorphic relation motivated by Finding 4:
\begin{definition}[MR3]
\label{def:mr3}
Given a program $P$ annotated with an annotation $a_1$ and another annotation $a_2$ that is equivalent to $a_1$, $P$ and $P_{a_1|a_2}$ should be analysis equivalent, 
i.e., $P\equiv P_{a_1|a_2}$, where $P_{a_1|a_2}$ denotes the resultant program produced by replacing annotation $a_1$ with $a_2$ in $P$.
\end{definition}

Our study shows that static analyzers sometimes fail to recognize all annotations with the same semantics. 
To address that limitation, we devise an equivalent annotation checker based on this metamorphic relation to automatically detect annotation-induced faults due to the root cause UEA.

\subsection{Annotated Program Generator}
\label{subsec:synthesis}

We design an annotated program generator to automatically derive annotated programs from the input programs.
The generated programs with annotations will be fed together with the original input programs to the static analyzers, 
and their analysis results will be checked by the checkers w.r.t.\ the aforementioned metamorphic relations.
The generator has three core components: 1) an annotation database, 2) an annotation generator, and 3) an annotation injector.

\subsubsection{Annotation Database} 
To build a database containing widely-used Java annotations,
we obtain annotations from two kinds of libraries in Maven Repo~\cite{mvncentral}: (1) the top 100 popular Java libraries and (2) the top 100 popular libraries labeled as ``Annotation libraries''. In total, our database contains 1616 annotations from 194 Java libraries (two libraries are duplicated, and four cannot be downloaded).

\subsubsection{Annotation Generator}

Our generator produces three types of annotations: (1) source level annotations, (2) dummy annotations, (3) equivalent annotation tuples. They correspond to the three checkers (i.e., ISC, ASC, and EAC).

\noindent\textbf{Source Level Annotations.}
\sys{} automatically selects source-level annotations from the database and generates annotation declarations without explicitly specified property values. 
Thus, \sys{} effectively associates all the annotations' properties to their default values.  

\noindent\textbf{Dummy Annotations.}
\sys uses the dummy annotation defined in Figure~\ref{fig:nosem}. 
We set its target to include all types of program elements that can be annotated so as to test the interplay between the annotation and static analyzers' AST-related operations more thoroughly.
We set its retention policy to RUNTIME so that the annotation will be retained for a longer time and hopefully can help us detect more annotation-induced faults at different stages of a static analyzer.
\vspace{-0.2cm}
\begin{figure}[H]
\begin{lstlisting}[style=javastyle]
import java.lang.annotation.*;
@Target({ElementType.METHOD, ...}) // Other targets omitted for space reasons
@Retention(RetentionPolicy.RUNTIME)
public @interface MockAnnotation {}
\end{lstlisting}
\caption{Definition of the dummy annotation.}
\label{fig:nosem}
\end{figure}
\vspace{-0.3cm}

\noindent\textbf{Equivalent Annotation Tuples.}
As explained in Section~\ref{sec:uea}, equivalent annotations should have similar semantics. \sys{} conservatively considers two annotations to be equivalent if and only if they have the same name and target set. As all the annotations in a tuple are semantics equivalent and added to identical program elements, their analysis scopes are the same. In total, we have collected 132 equivalent annotation tuples. If \sys{} were to use all 132 tuples, too many mutants could be generated (as each tuple leads to at least two annotated programs being generated). Hence, we select 24 tuples based on the top 30 AIF prone annotations identified in RQ1. All tuples have been manually verified that they are indeed equivalent tuples by two authors. Additionally, using all tuples will significantly increase running time, e.g., for PMD, the fastest among evaluated tools, tuple selection can reduce running time by 91.3\% (all tuples = 69 hours, selected tuples = 6 hours) while finding the same number of bugs.

\subsubsection{Annotation Injector} Given an input program $P$ and an annotation $a$ generated by the annotation generator,
the annotation injector first analyzes the annotation to determine the set of valid targets for it, then goes through the program to collect specific locations where the annotation can be applied, and finally automatically inserts the annotation in all those locations.
For example, if ElementType.METHOD is a valid target for an annotation, the annotation can be applied to annotate method declarations.
The output of the annotation injector is a set of $P$'s variants, or mutants, each with the annotation being injected in a different location. Some injected annotations may cause compilation errors (around 2\%) if their corresponding properties require explicit initialization, so \sys{} discards these syntactically invalid variants before proceeding to subsequent steps.

\section{Effectiveness of \sys{}}

We applied \sys{} to \pmd{}, \sbugs{}, \cstyle{}, \infer{}, \sonar{} and \soot{} and conducted experiments to measure the effectiveness of \sys{} by reusing test suites from the official repositories of static analyzer as the seed corpus as prior work shows that these tests can help us cover more rule checkers to reveal more faults~\cite{statfier}.
For static analyzers that require compilation (e.g., SpotBugs), we compile each program using Oracle JDK 17. All experiments were conducted on a machine with Intel Xeon(R) 6134 CPU 3.20GHz and 192GB RAM.
For each checker and its corresponding annotations, we run \sys{} on all analyzers in parallel until all generated mutants have been evaluated and do not set any timeout. We did not test \sys{} on known issues as it was designed based on insights gained from these issues. Testing \sys{} on the same issues would introduce bias. We also identify two challenges in evaluating \sys{} on known issues:
(1) it involves building old versions of analyzers from their source code, which can be quite demanding (e.g., due to the absence of required external libraries and the intricacies of the compilation process),
(2) we are missing compilable input programs to reproduce some known issues, but those programs can be hard to construct manually, and \sys{} requires them as the input.

\input{tab-NewExperiment}

Table~\ref{tab:new_experiment} shows the experiment results.
We measure the effectiveness of \sys{} by counting the unique faults detected by each checker (``\#UniqFaults'' column). Specifically, we manually analyze the root causes of the identified faults and remove duplicated ones.
Notably, we consider two faults duplicated if they are in (1) an identical rule checker and (2) an identical faulty location (determined by root cause diagnosis) in a static analyzer.
Table~\ref{tab:new_experiment} shows that \sys{} found \faultcnt{} bugs in evaluated static analyzers,
and \fixedfaultcnt{} have been fixed via merged pull requests (9 by developers and 13 by authors).
Overall, \sys{} finds the greatest number of faults using ISC.
This result is consistent with Finding 2, which shows the prevalence of IS in static analyzers.
The ``Time'' column shows the minimum and maximum total execution time for all checkers on the six static analyzers. Although different checkers use the same seed corpus,
the time taken by different checkers varies because the number of annotations and the number of valid program locations to inject these annotations are different.

\noindent\textbf{Fix Strategies for \sys{}'s Found Bugs.} 
To further analyze the fixed issues, we classify the fix strategies of the \fixedfaultcnt{} fixed issues.
All of them fit into our taxonomy of fix strategies: 14 by FAF, 3 by FGI, 2 by FAN, and 1 by FIA.
The result \emph{illustrates the generality of our taxonomy}.

\noindent\textbf{Limitations.} Like other testing tools, \sys{} also reports \fpcnt{} FPs (``\#FP'' column in Table~\ref{tab:new_experiment}).
Our manual analysis of the FPs revealed that all FPs are caused by the source code changes induced by applying $MR1$ to recover annotation semantics.
For example, \textit{@NoArgsConstructor} is semantically equivalent to injecting a no-argument constructor into source code, but the constructor triggers a \textit{UnnecessaryConstructor} warning in \pmd{}, causing an FP (this extra warning misleads \sys{} into thinking that the programs before and after annotation processing are not analysis equivalent). Another limitation is \sys{} requires manual effort to verify the correctness of the 43 identified unique faults.

\subsection{Case Study}

We select three faults found by \sys{} to show \sys{}'s fault finding capability. For each fault, we present its root cause, the affected analyzer, and how \sys{} found the issue.

\noindent \textbf{A crash in \pmd{}~\cite{pmd4152c1}.} Finding 5 shows that static analyzers cannot handle special annotation syntax as developers tend to neglect them. Figure~\ref{fig:crashcase} shows an example of crash in \pmd{} discovered by \sys{}. At line 7 of this example, \pmd{} fails to process the annotation \textit{@DummyAnnotation} placed on the class constructor reference \textit{::new} due to the grammar issue, consequently leading to a runtime crash. The developers have fixed this issue upon receiving our report.

\begin{figure}[!htbp]
\begin{lstlisting}[style=javastyle]
import java.util.function.Function;
public class Main {
  public class Inner {
    public Inner(Object o) {}
  }
  public Function func(Main this) {
    return @DummyAnnotation Main.Inner::new; // Crash
  }
}
\end{lstlisting}
\caption{A crash example in \pmd{} detected by \sys{}}
\label{fig:crashcase}
\end{figure}

\noindent \textbf{An FP in \sonar{}~\cite{sonarfp}.} Figure~\ref{fig:fpcase1} shows a fault caused by incomplete semantics. \sonar{} reports an unclosed stream warning at line 2, but it is an FP because the \textit{@Cleanup} annotation will generate a \textit{try-finally} statement to close \textit{FileInputStream} in the \textit{finally} block. This issue has been confirmed and marked as ``Major'' priority by the developer, indicating the importance of the fault.

\begin{figure}[!htbp]
\begin{lstlisting}[style=javastyle]
public static void main(String[] args) throws IOException {
  @Cleanup InputStream in = new FileInputStream(args[0]); // FP
  ...
\end{lstlisting}
\caption{An FP example in \sonar{} detected by \sys{}}
\label{fig:fpcase1}
\end{figure}

\noindent \textbf{An FP in \pmd{}~\cite{pmd4487}.} 
Figure~\ref{pmdfpcase} shows that \pmd{} reported a warning against the unnecessary constructor at line 4.
But that is an FP because the annotation \textit{@Inject} uses this constructor for dependency injection. 
\pmd{} does not consider the annotation ``com.google.inject.Inject'' in Figure~\ref{pmdfpcase}, but it has considered another equivalent annotation ``javax.inject.Inject''.
EAC can automatically detect this FP. We have fixed this bug via a merged PR in collaboration with developers.

\begin{figure}[!htbp]
\begin{lstlisting}[style=javastyle]
import com.google.inject.Inject;
public class Foo { private Foo() {} }
public class Bar extends Foo {
  @Inject public Bar() {} // Report a warning, but it is an FP
}
\end{lstlisting}
\caption{An FP example in \pmd{} detected by \sys{}}
\label{pmdfpcase}
\end{figure}

%% file: tab-NewExperiment.tex
\begin{table}[!htbp]
\small
\caption{Effectiveness of \sys{}}
\label{tab:new_experiment}
\centering
\begin{tabular*}{\linewidth}{@{\extracolsep{\fill} }lrrrrr}
\toprule
Checker & \#Violations & \#UniqFaults & \#FP & \#Fixed & Time (min, max) (hour) \\
\midrule
ISC & 258 & 19 & 8 & 11 & (4,62) \\
ASC & 52 & 8 & 0 & 4 & (2,24) \\
EAC & 123 & 16 & 0 & 5 & (6,87) \\
\hline
Overall & 433 & 43 & 8 & 20 & (6,87) \\
\bottomrule
\end{tabular*}
\end{table}

%% file: 9-Implication.tex
\section{Implication}

We discuss below the implications for developers and researchers based on our study findings:

\noindent \textbf{Implication for Developers.}
Our study identifies the common root causes and their corresponding symptoms and fix strategies that may help developers of static analyzers to detect, understand, and repair faults caused by annotation. We also study AIF prone annotations, implying that developers should pay attention to these annotations (Finding 1). Based on this finding, we design \sys{} to select AIF prone annotations. In the future, it is worthwhile to investigate more advanced techniques for annotation selection. Based on the two most common root causes of annotation-induced faults in our study (IS and IAT), we realized that developers of static analyzers tend to either (1) be unaware of the semantics encoded by annotations (Finding 2) or (2) neglect the impact of annotations on program ASTs (Finding 3). Hence, we hope that our study will raise awareness among developers on the impacts of Java annotations on static analyzers to improve the accuracy and correctness of static analyzers. In terms of the static analyzer workflow, our study revealed that \emph{developers should pay careful attention to annotations when performing syntax analysis} because all studied static analyzers have annotation-induced faults in the syntactic analysis stage, especially when annotations are placed on the types such as generic type arguments and type casts since JSR-308~\cite{jsr308}. With the evolution of Java specification, developers should also consider annotation-induced faults when updating the grammar (Finding 5). Meanwhile, as our study also revealed that there exists a set of equivalent annotations that come from different libraries or different versions of the same libraries (Finding 4), \emph{developers should consider these related annotation libraries when designing rule checkers to provide comprehensive support for the related annotations}. 

\noindent \textbf{Implication for Researchers.}
Our study and proposed framework lay the foundation for research in three promising directions. 
First, \textit{encoding the semantics of annotations into static analyzers is essential in improving the accuracy of the analysis} because current analyzers fail to model the behavior of annotations well.
Incomplete semantics is the most common root cause of AIFs in our study (Finding 2) and the greatest number of bugs detected by \sys{}. Therefore, failing to solve this problem can affect the fault detection capability of static analyzers. Second, \emph{detecting AIFs is necessary but yet often neglected}. For static analyzers, eliminating FPs is a worthwhile and long-term research direction~\cite{deepinfer2022icse}. As shown in Table~\ref{tab:symptomdistribution}, FP is the most common symptom caused by AIFs. Consequently, detecting AIFs is rewarding for reducing FPs. Metamorphic testing is a promising approach for this purpose. Researchers can produce annotated program pairs and compare their analysis reports to detect FPs (such as ISC). Third, \emph{our fix strategies (Finding 8--10) serve as preliminary studies for future research on automated repair of AIFs}. We observe that several fix strategies in our study can be automated to reduce the effort in fixing them (e.g., Fix Annotation Filter (FAF) can fix more than half of the issues). Most of them are implemented by creating an annotation filter or extending an existing filter.

%% file: 10-Threat.tex
\section{Threats to Validity}

\noindent\textbf{External.} Our study may not generalize beyond the studied analyzers and other programming languages beyond Java. To ensure the generalizability of our findings, we select \numtool{} representative static analyzers (Section~\ref{sec:toolselection}), and we systematically analyze \issuecnt{} issues in these analyzers.
The selection of whole-program analyzers may also have potential implications for the generalizability of the results. As stated in Section~\ref{sec:toolselection}, we only consider static analyzers from open-source repositories with over 2,000 stars, solely choosing Soot as a representative of whole-program analyzers. However, our study shows \sys{} can find bugs in lightweight tools (e.g., PMD), and Soot shares many root causes, symptoms, and fix strategies with lightweight tools. We leave a more comprehensive study on lightweight and whole-program static analyzers for future work.

\noindent\textbf{Internal.} Manually labeling annotation-induced faults may be subjective and biased. To reduce this threat, we refer to previous taxonomies~\cite{sun2017mlbug, zhang2018tfbug, chen2019cloudexception, shen2021dlbugstudy, zhang2020dljob} and adopt an open-coding scheme to adapt taxonomies to annotation-induced faults. Our code may have bugs that can affect the evaluation results. To mitigate this threat, we have made our tool and data publicly available.

%% file: 11-RelatedWork.tex
\section{Related Work}

\noindent\textbf{Studies on Annotation.} Several studies have investigated the common code annotation practices in Java~\cite{tse2019annotation,pinheiro2020mutating,nuryyev2022mining,rocha2011annotations,parnin2013adoption}. These studies showed that annotations are widely adopted by Java developers and served as motivations for our study.
Among these studies, the study of annotation-related faults and the mutation operators that mimic these faults is the closest related work~\cite{pinheiro2020mutating} (e.g., the insertion of annotations in \sys{} is similar to the ADA operator that adds an annotation to a valid target in prior work, but the ADA operator requires users to manually specify the annotation whereas we generate annotation from our annotation database~\cite{pinheiro2020mutating}). Nevertheless, our study and our proposed technique are different from prior studies in several important aspects: (1) prior study only identified two categories of annotation-related faults (``misuse'' and ``wrong annotation parsing''), whereas our study focuses on the impacts on annotation-induced faults in static analyzers (the root cause categories of annotation-induced faults in our study is more diverse than that in prior study); (2) prior technique focuses on mutating code annotations~\cite{pinheiro2020mutating} via a set of operators, but it cannot detect annotation-induced faults due to the lack of oracle, whereas \sys{} injects annotations into input programs and uses a well-designed set of oracles to find annotated-induced faults in static analyzers. As the annotation-induced faults identified in our study are more diverse, these faults may affect other types of software applications (beyond static analyzers).
In the future, it is worthwhile to study annotation-induced faults in a broader domain.

\noindent\textbf{Studies on Faults.} There are several studies on faults~\cite{shen2021dlbugstudy, issta2022canalyzer, issta2016gccstudy, zhang2018tfbug, issre2012mlsystem, msr2017blockchain}.
Although prior studies investigated faults for diverse types of software systems (e.g., machine learning systems~\cite{shen2021dlbugstudy, issre2012mlsystem}, blockchains~\cite{msr2017blockchain}, compilers~\cite{issta2016gccstudy}, and static analyzers~\cite{issta2022canalyzer}), they did not cover AIFs of static analyzers. The most closely related to our study is the recent study~\cite{issta2022canalyzer} that evaluates the vulnerability detection capability of C static analyzers. Our work differs from the prior study in several aspects: (1) we conduct the first study to understand annotation-induced faults of static analyzers, covering the root cause, symptom, fix strategy, and AIF prone annotations; and (2) we propose metamorphic relations to detect annotation-induced faults, and implement a testing framework.
A recent study~\cite{issta2023sast} built benchmarks and evaluated the effectiveness and performance of static analyzers but did not investigate AIFs fault and failed to detect AIFs.

\noindent\textbf{Testing Static Analyzers.} Several techniques have been proposed for testing analyzers~\cite{2022icpcfinder, cuoq2012testing, klinger2019differentially, chen2018metamorphic, araujo2016correlating, ase2021npe, icse2023soottesting, li2024tracking}.
Some use equivalent relations as the metamorphic relation to solve the lack of oracle problem in static analyzer testing~\cite{le2014compiler, fse2021datalog, icse2016testanalyzer,statfier}.
Similarly, we use equivalent relation (i.e., constructing three behavior equivalent properties and checking for differential analysis results of equivalent mutants) to address the oracle problem, but our equivalent properties have incorporated the characteristics of annotations to find annotation-induced faults.
While prior approaches~\cite{ase2021npe, icse2023soottesting,statfier} can find other types of bugs in static analyzers, none of them focuses on  AIF faults (and they could not generate annotated programs as in \sys{} to detect AIFs).

%% file: 12-Conclusion.tex
\section{Conclusion}
We conduct the first comprehensive study which focuses on understanding and detecting annotation-induced faults of static analyzers as annotation has become a popular programming paradigm. We manually investigate \issuecnt{} issues from six representative and diverse static analyzers (\sonar{}, \cstyle{}, \pmd{}, \sbugs{}, \infer{}, and \soot{}), identify \rootcausecnt{} root causes, \symptomcnt{} symptoms, and \fixstrategycnt{} fix strategies. Moreover, we also summarize \findingcnt{} findings. Based on these findings, we introduce a set of guidelines for AIFs detection and repair, and propose the first metamorphic testing based framework \sys{} to automatically find AIFs in static analyzers. With our annotation synthesizer and three metamorphic relations, it can generate new tests based on official test suites and find \faultcnt{} faults where \fixedfaultcnt{} of them have been fixed.

\section*{Data Availability} 
The experimental data and source code are available at: https://annaresearch.github.io/.

%% file: 0-Main.bbl

\begin{thebibliography}{85}


\ifx \showCODEN    \undefined \def \showCODEN     #1{\unskip}     \fi
\ifx \showDOI      \undefined \def \showDOI       #1{#1}\fi
\ifx \showISBNx    \undefined \def \showISBNx     #1{\unskip}     \fi
\ifx \showISBNxiii \undefined \def \showISBNxiii  #1{\unskip}     \fi
\ifx \showISSN     \undefined \def \showISSN      #1{\unskip}     \fi
\ifx \showLCCN     \undefined \def \showLCCN      #1{\unskip}     \fi
\ifx \shownote     \undefined \def \shownote      #1{#1}          \fi
\ifx \showarticletitle \undefined \def \showarticletitle #1{#1}   \fi
\ifx \showURL      \undefined \def \showURL       {\relax}        \fi
\providecommand\bibfield[2]{#2}
\providecommand\bibinfo[2]{#2}
\providecommand\natexlab[1]{#1}
\providecommand\showeprint[2][]{arXiv:#2}

\bibitem[adangel(2019)]%
        {pmd1782}
\bibfield{author}{\bibinfo{person}{adangel}.} \bibinfo{year}{2019}\natexlab{}.
\newblock \bibinfo{booktitle}{\emph{NoPackage: False Negative for enums}}.
\newblock
\urldef\tempurl%
\url{https://github.com/pmd/pmd/issues/1782}
\showURL{%
\tempurl}


\bibitem[Alhadidi et~al\mbox{.}(2009)]%
        {aosd2009dataflow}
\bibfield{author}{\bibinfo{person}{Dima Alhadidi}, \bibinfo{person}{Amine Boukhtouta}, \bibinfo{person}{Nadia Belblidia}, \bibinfo{person}{Mourad Debbabi}, {and} \bibinfo{person}{Prabir Bhattacharya}.} \bibinfo{year}{2009}\natexlab{}.
\newblock \showarticletitle{The dataflow pointcut: a formal and practical framework}. In \bibinfo{booktitle}{\emph{Proceedings of the 8th ACM international conference on Aspect-oriented software development}}. \bibinfo{pages}{15--26}.
\newblock


\bibitem[Ara{\'u}jo et~al\mbox{.}(2016)]%
        {araujo2016correlating}
\bibfield{author}{\bibinfo{person}{Cl{\'a}udio~A Ara{\'u}jo}, \bibinfo{person}{Marcio~E Delamaro}, \bibinfo{person}{Jos{\'e}~C Maldonado}, {and} \bibinfo{person}{Auri~MR Vincenzi}.} \bibinfo{year}{2016}\natexlab{}.
\newblock \showarticletitle{Correlating automatic static analysis and mutation testing: towards incremental strategies}.
\newblock \bibinfo{journal}{\emph{Journal of Software Engineering Research and Development}} \bibinfo{volume}{4}, \bibinfo{number}{1} (\bibinfo{year}{2016}), \bibinfo{pages}{1--32}.
\newblock


\bibitem[Ayewah and Pugh(2010)]%
        {fixit}
\bibfield{author}{\bibinfo{person}{Nathaniel Ayewah} {and} \bibinfo{person}{William Pugh}.} \bibinfo{year}{2010}\natexlab{}.
\newblock \showarticletitle{The Google FindBugs fixit}. \bibinfo{pages}{241--252}.
\newblock
\urldef\tempurl%
\url{https://doi.org/10.1145/1831708.1831738}
\showDOI{\tempurl}


\bibitem[Ayewah et~al\mbox{.}(2007)]%
        {ayewah2007evaluating}
\bibfield{author}{\bibinfo{person}{Nathaniel Ayewah}, \bibinfo{person}{William Pugh}, \bibinfo{person}{J~David Morgenthaler}, \bibinfo{person}{John Penix}, {and} \bibinfo{person}{YuQian Zhou}.} \bibinfo{year}{2007}\natexlab{}.
\newblock \showarticletitle{Evaluating static analysis defect warnings on production software}. In \bibinfo{booktitle}{\emph{Proceedings of the 7th ACM SIGPLAN-SIGSOFT workshop on Program analysis for software tools and engineering}}. \bibinfo{pages}{1--8}.
\newblock


\bibitem[Belle(2022)]%
        {sonarfp}
\bibfield{author}{\bibinfo{person}{Belle}.} \bibinfo{year}{2022}\natexlab{}.
\newblock \bibinfo{booktitle}{\emph{A false positive about rule RSPEC-2095}}.
\newblock
\urldef\tempurl%
\url{https://community.sonarsource.com/t/a-false-positive-about-rule-rspec-2095/67536}
\showURL{%
\tempurl}


\bibitem[Belle-PL(2022)]%
        {pmd4152c1}
\bibfield{author}{\bibinfo{person}{Belle-PL}.} \bibinfo{year}{2022}\natexlab{}.
\newblock \bibinfo{booktitle}{\emph{Parse error on array type annotations}}.
\newblock
\urldef\tempurl%
\url{https://github.com/pmd/pmd/issues/4152#issuecomment-1277447394}
\showURL{%
\tempurl}


\bibitem[Cadar and Donaldson(2016)]%
        {icse2016testanalyzer}
\bibfield{author}{\bibinfo{person}{Cristian Cadar} {and} \bibinfo{person}{Alastair~F Donaldson}.} \bibinfo{year}{2016}\natexlab{}.
\newblock \showarticletitle{Analysing the program analyser}. In \bibinfo{booktitle}{\emph{Proceedings of the 38th International Conference on Software Engineering Companion}}. \bibinfo{pages}{765--768}.
\newblock


\bibitem[Campbell and Papapetrou(2013)]%
        {sonarinaction}
\bibfield{author}{\bibinfo{person}{G~Ann Campbell} {and} \bibinfo{person}{Patroklos~P Papapetrou}.} \bibinfo{year}{2013}\natexlab{}.
\newblock \bibinfo{booktitle}{\emph{SonarQube in action}}.
\newblock \bibinfo{publisher}{Manning Publications Co.}
\newblock


\bibitem[Cazzola and Vacchi(2014)]%
        {clss2014java}
\bibfield{author}{\bibinfo{person}{Walter Cazzola} {and} \bibinfo{person}{Edoardo Vacchi}.} \bibinfo{year}{2014}\natexlab{}.
\newblock \showarticletitle{@ Java: Bringing a richer annotation model to Java}.
\newblock \bibinfo{journal}{\emph{Computer Languages, Systems \& Structures}} \bibinfo{volume}{40}, \bibinfo{number}{1} (\bibinfo{year}{2014}), \bibinfo{pages}{2--18}.
\newblock


\bibitem[CheckStyle(2022)]%
        {checkstylelink}
\bibfield{author}{\bibinfo{person}{CheckStyle}.} \bibinfo{year}{2022}\natexlab{}.
\newblock \bibinfo{booktitle}{\emph{Checkstyle}}.
\newblock
\urldef\tempurl%
\url{https://checkstyle.sourceforge.io/}
\showURL{%
\tempurl}


\bibitem[Chen et~al\mbox{.}(2019)]%
        {chen2019cloudexception}
\bibfield{author}{\bibinfo{person}{Haicheng Chen}, \bibinfo{person}{Wensheng Dou}, \bibinfo{person}{Yanyan Jiang}, {and} \bibinfo{person}{Feng Qin}.} \bibinfo{year}{2019}\natexlab{}.
\newblock \showarticletitle{Understanding exception-related bugs in large-scale cloud systems}. In \bibinfo{booktitle}{\emph{2019 34th IEEE/ACM International Conference on Automated Software Engineering (ASE)}}. IEEE, \bibinfo{pages}{339--351}.
\newblock


\bibitem[Chen et~al\mbox{.}(2018)]%
        {chen2018metamorphic}
\bibfield{author}{\bibinfo{person}{Tsong~Yueh Chen}, \bibinfo{person}{Fei-Ching Kuo}, \bibinfo{person}{Huai Liu}, \bibinfo{person}{Pak-Lok Poon}, \bibinfo{person}{Dave Towey}, \bibinfo{person}{TH Tse}, {and} \bibinfo{person}{Zhi~Quan Zhou}.} \bibinfo{year}{2018}\natexlab{}.
\newblock \showarticletitle{Metamorphic testing: A review of challenges and opportunities}.
\newblock \bibinfo{journal}{\emph{ACM Computing Surveys (CSUR)}} \bibinfo{volume}{51}, \bibinfo{number}{1} (\bibinfo{year}{2018}), \bibinfo{pages}{1--27}.
\newblock


\bibitem[Christakis and Bird(2016)]%
        {christakis2016developers}
\bibfield{author}{\bibinfo{person}{Maria Christakis} {and} \bibinfo{person}{Christian Bird}.} \bibinfo{year}{2016}\natexlab{}.
\newblock \showarticletitle{What developers want and need from program analysis: an empirical study}. In \bibinfo{booktitle}{\emph{Proceedings of the 31st IEEE/ACM international conference on automated software engineering}}. \bibinfo{pages}{332--343}.
\newblock


\bibitem[Cuoq et~al\mbox{.}(2012)]%
        {cuoq2012testing}
\bibfield{author}{\bibinfo{person}{Pascal Cuoq}, \bibinfo{person}{Benjamin Monate}, \bibinfo{person}{Anne Pacalet}, \bibinfo{person}{Virgile Prevosto}, \bibinfo{person}{John Regehr}, \bibinfo{person}{Boris Yakobowski}, {and} \bibinfo{person}{Xuejun Yang}.} \bibinfo{year}{2012}\natexlab{}.
\newblock \showarticletitle{Testing static analyzers with randomly generated programs}. In \bibinfo{booktitle}{\emph{NASA Formal Methods Symposium}}. Springer, \bibinfo{pages}{120--125}.
\newblock


\bibitem[Czarnecki and Eisenecker(2000)]%
        {CzarneckiEisenecker00}
\bibfield{author}{\bibinfo{person}{Krysztof Czarnecki} {and} \bibinfo{person}{Ulrich Eisenecker}.} \bibinfo{year}{2000}\natexlab{}.
\newblock \bibinfo{booktitle}{\emph{Generative Programming: Methods, Tools, and Applications}}.
\newblock \bibinfo{publisher}{Addison-Wesley}, \bibinfo{address}{Boston, MA}.
\newblock
\showISBNx{978-0-201-30977-5}


\bibitem[Do et~al\mbox{.}(2022)]%
        {tse2022usestaticanalysis}
\bibfield{author}{\bibinfo{person}{Lisa Nguyen~Quang Do}, \bibinfo{person}{James~R. Wright}, {and} \bibinfo{person}{Karim Ali}.} \bibinfo{year}{2022}\natexlab{}.
\newblock \showarticletitle{Why Do Software Developers Use Static Analysis Tools? A User-Centered Study of Developer Needs and Motivations}.
\newblock \bibinfo{journal}{\emph{IEEE Transactions on Software Engineering}} \bibinfo{volume}{48}, \bibinfo{number}{3} (\bibinfo{year}{2022}), \bibinfo{pages}{835--847}.
\newblock
\urldef\tempurl%
\url{https://doi.org/10.1109/TSE.2020.3004525}
\showDOI{\tempurl}


\bibitem[fluxroot(2019)]%
        {pmd1641}
\bibfield{author}{\bibinfo{person}{fluxroot}.} \bibinfo{year}{2019}\natexlab{}.
\newblock \bibinfo{booktitle}{\emph{False-positive with Lombok and inner classes}}.
\newblock
\urldef\tempurl%
\url{https://github.com/pmd/pmd/issues/1641}
\showURL{%
\tempurl}


\bibitem[Gumowski(2015)]%
        {sonar1167}
\bibfield{author}{\bibinfo{person}{Michael Gumowski}.} \bibinfo{year}{2015}\natexlab{}.
\newblock \bibinfo{booktitle}{\emph{Annotations should be handled in all cases allowed by java 8}}.
\newblock
\urldef\tempurl%
\url{https://sonarsource.atlassian.net/browse/SONARJAVA-1167}
\showURL{%
\tempurl}


\bibitem[Gumowski(2016)]%
        {sonar1513}
\bibfield{author}{\bibinfo{person}{Michael Gumowski}.} \bibinfo{year}{2016}\natexlab{}.
\newblock \bibinfo{booktitle}{\emph{Classes annotated with lombok's @EqualsAndHashCode should be ignored}}.
\newblock
\urldef\tempurl%
\url{https://sonarsource.atlassian.net/browse/SONARJAVA-1513}
\showURL{%
\tempurl}


\bibitem[Gumowski(2017a)]%
        {sonar2205}
\bibfield{author}{\bibinfo{person}{Michael Gumowski}.} \bibinfo{year}{2017}\natexlab{a}.
\newblock \bibinfo{booktitle}{\emph{Annotations on type in a fully qualified name are not resolved}}.
\newblock
\urldef\tempurl%
\url{https://sonarsource.atlassian.net/browse/SONARJAVA-2205}
\showURL{%
\tempurl}


\bibitem[Gumowski(2017b)]%
        {sonar2083}
\bibfield{author}{\bibinfo{person}{Michael Gumowski}.} \bibinfo{year}{2017}\natexlab{b}.
\newblock \bibinfo{booktitle}{\emph{FP on S1128 (UselessImportCheck) when annotation is used in fully qualified name}}.
\newblock
\urldef\tempurl%
\url{https://sonarsource.atlassian.net/browse/SONARJAVA-2083}
\showURL{%
\tempurl}


\bibitem[Gumowski(2019a)]%
        {sonar3045}
\bibfield{author}{\bibinfo{person}{Michael Gumowski}.} \bibinfo{year}{2019}\natexlab{a}.
\newblock \bibinfo{booktitle}{\emph{Annotations are not always resolved when used in parameterized types}}.
\newblock
\urldef\tempurl%
\url{https://sonarsource.atlassian.net/browse/SONARJAVA-3045}
\showURL{%
\tempurl}


\bibitem[Gumowski(2019b)]%
        {sonar3108}
\bibfield{author}{\bibinfo{person}{Michael Gumowski}.} \bibinfo{year}{2019}\natexlab{b}.
\newblock \bibinfo{booktitle}{\emph{Generating starting states make analysis crash (OutOfMemory) when too many annotated parameters}}.
\newblock
\urldef\tempurl%
\url{https://sonarsource.atlassian.net/browse/SONARJAVA-3108}
\showURL{%
\tempurl}


\bibitem[{Habib} and {Pradel}(2018)]%
        {how_many}
\bibfield{author}{\bibinfo{person}{A. {Habib}} {and} \bibinfo{person}{M. {Pradel}}.} \bibinfo{year}{2018}\natexlab{}.
\newblock \showarticletitle{How Many of All Bugs Do We Find? {A} Study of Static Bug Detectors}. In \bibinfo{booktitle}{\emph{2018 33rd IEEE/ACM International Conference on Automated Software Engineering (ASE)}}. \bibinfo{pages}{317--328}.
\newblock
\urldef\tempurl%
\url{https://doi.org/10.1145/3238147.3238213}
\showDOI{\tempurl}


\bibitem[Infer(2022)]%
        {inferlink}
\bibfield{author}{\bibinfo{person}{Infer}.} \bibinfo{year}{2022}\natexlab{}.
\newblock \bibinfo{booktitle}{\emph{Infer Static Analyzer}}.
\newblock
\urldef\tempurl%
\url{https://fbinfer.com/}
\showURL{%
\tempurl}


\bibitem[Jaquier(2019)]%
        {sonar3174}
\bibfield{author}{\bibinfo{person}{Quentin Jaquier}.} \bibinfo{year}{2019}\natexlab{}.
\newblock \bibinfo{booktitle}{\emph{Support Java 11 Generated annotation}}.
\newblock
\urldef\tempurl%
\url{https://sonarsource.atlassian.net/browse/SONARJAVA-3174}
\showURL{%
\tempurl}


\bibitem[Jaquier(2020)]%
        {sonar3536}
\bibfield{author}{\bibinfo{person}{Quentin Jaquier}.} \bibinfo{year}{2020}\natexlab{}.
\newblock \bibinfo{booktitle}{\emph{Consistently support Nullable/CheckForNull/Nonnull annotations in rules}}.
\newblock
\urldef\tempurl%
\url{https://sonarsource.atlassian.net/browse/SONARJAVA-3536}
\showURL{%
\tempurl}


\bibitem[Jaquier(2023)]%
        {sonar3438}
\bibfield{author}{\bibinfo{person}{Quentin Jaquier}.} \bibinfo{year}{2023}\natexlab{}.
\newblock \bibinfo{booktitle}{\emph{S5122: ClassCastException when annotation is defined with an identifier}}.
\newblock
\urldef\tempurl%
\url{https://sonarsource.atlassian.net/browse/SONARJAVA-3438}
\showURL{%
\tempurl}


\bibitem[Johnson et~al\mbox{.}(2013)]%
        {johnson2013don}
\bibfield{author}{\bibinfo{person}{Brittany Johnson}, \bibinfo{person}{Yoonki Song}, \bibinfo{person}{Emerson Murphy-Hill}, {and} \bibinfo{person}{Robert Bowdidge}.} \bibinfo{year}{2013}\natexlab{}.
\newblock \showarticletitle{Why don't software developers use static analysis tools to find bugs?}. In \bibinfo{booktitle}{\emph{2013 35th International Conference on Software Engineering (ICSE)}}. IEEE, \bibinfo{pages}{672--681}.
\newblock


\bibitem[Juneau(2022)]%
        {jsr308}
\bibfield{author}{\bibinfo{person}{Josh Juneau}.} \bibinfo{year}{2022}\natexlab{}.
\newblock \bibinfo{booktitle}{\emph{JSR 308 Explained: Java Type Annotations}}.
\newblock
\urldef\tempurl%
\url{https://www.oracle.com/technical-resources/articles/java/ma14-architect-annotations.html}
\showURL{%
\tempurl}


\bibitem[Kharkar et~al\mbox{.}(2022)]%
        {deepinfer2022icse}
\bibfield{author}{\bibinfo{person}{Anant Kharkar}, \bibinfo{person}{Roshanak~Zilouchian Moghaddam}, \bibinfo{person}{Matthew Jin}, \bibinfo{person}{Xiaoyu Liu}, \bibinfo{person}{Xin Shi}, \bibinfo{person}{Colin Clement}, {and} \bibinfo{person}{Neel Sundaresan}.} \bibinfo{year}{2022}\natexlab{}.
\newblock \showarticletitle{Learning to reduce false positives in analytic bug detectors}. In \bibinfo{booktitle}{\emph{Proceedings of the 44th International Conference on Software Engineering}}. \bibinfo{pages}{1307--1316}.
\newblock


\bibitem[Kim et~al\mbox{.}(2021)]%
        {icse2021annotationmaintenance}
\bibfield{author}{\bibinfo{person}{Dong~Jae Kim}, \bibinfo{person}{Nikolaos Tsantalis}, \bibinfo{person}{Tse-Hsun Chen}, {and} \bibinfo{person}{Jinqiu Yang}.} \bibinfo{year}{2021}\natexlab{}.
\newblock \showarticletitle{Studying Test Annotation Maintenance in the Wild}. In \bibinfo{booktitle}{\emph{2021 IEEE/ACM 43rd International Conference on Software Engineering (ICSE)}}. \bibinfo{pages}{62--73}.
\newblock
\urldef\tempurl%
\url{https://doi.org/10.1109/ICSE43902.2021.00019}
\showDOI{\tempurl}


\bibitem[Klinger et~al\mbox{.}(2019)]%
        {klinger2019differentially}
\bibfield{author}{\bibinfo{person}{Christian Klinger}, \bibinfo{person}{Maria Christakis}, {and} \bibinfo{person}{Valentin W{\"u}stholz}.} \bibinfo{year}{2019}\natexlab{}.
\newblock \showarticletitle{Differentially testing soundness and precision of program analyzers}. In \bibinfo{booktitle}{\emph{Proceedings of the 28th ACM SIGSOFT International Symposium on Software Testing and Analysis}}. \bibinfo{pages}{239--250}.
\newblock


\bibitem[konrad jamrozik(2013)]%
        {soot123}
\bibfield{author}{\bibinfo{person}{konrad jamrozik}.} \bibinfo{year}{2013}\natexlab{}.
\newblock \bibinfo{booktitle}{\emph{Regression: ERROR/dalvikvm(1854): Invalid type descriptor: 'dalvik.annotation.EnclosingClass'}}.
\newblock
\urldef\tempurl%
\url{https://github.com/soot-oss/soot/issues/123}
\showURL{%
\tempurl}


\bibitem[krzyk(2015)]%
        {cstyle2202}
\bibfield{author}{\bibinfo{person}{krzyk}.} \bibinfo{year}{2015}\natexlab{}.
\newblock \bibinfo{booktitle}{\emph{SuppressWarnings should support CamelCase}}.
\newblock
\urldef\tempurl%
\url{https://github.com/checkstyle/checkstyle/issues/2202}
\showURL{%
\tempurl}


\bibitem[Le et~al\mbox{.}(2014)]%
        {le2014compiler}
\bibfield{author}{\bibinfo{person}{Vu Le}, \bibinfo{person}{Mehrdad Afshari}, {and} \bibinfo{person}{Zhendong Su}.} \bibinfo{year}{2014}\natexlab{}.
\newblock \showarticletitle{Compiler validation via equivalence modulo inputs}.
\newblock \bibinfo{journal}{\emph{ACM Sigplan Notices}} \bibinfo{volume}{49}, \bibinfo{number}{6} (\bibinfo{year}{2014}), \bibinfo{pages}{216--226}.
\newblock


\bibitem[lgoldstein(2017)]%
        {cstyle4472}
\bibfield{author}{\bibinfo{person}{lgoldstein}.} \bibinfo{year}{2017}\natexlab{}.
\newblock \bibinfo{booktitle}{\emph{EmptyBlock: NPE on annotation declaration}}.
\newblock
\urldef\tempurl%
\url{https://github.com/checkstyle/checkstyle/issues/4472}
\showURL{%
\tempurl}


\bibitem[Li and Yang(2024)]%
        {li2024tracking}
\bibfield{author}{\bibinfo{person}{Junjie Li} {and} \bibinfo{person}{Jinqiu Yang}.} \bibinfo{year}{2024}\natexlab{}.
\newblock \showarticletitle{Tracking the Evolution of Static Code Warnings: the State-of-the-Art and a Better Approach}.
\newblock \bibinfo{journal}{\emph{IEEE Transactions on Software Engineering}} (\bibinfo{year}{2024}).
\newblock


\bibitem[Lipp et~al\mbox{.}(2022)]%
        {issta2022canalyzer}
\bibfield{author}{\bibinfo{person}{Stephan Lipp}, \bibinfo{person}{Sebastian Banescu}, {and} \bibinfo{person}{Alexander Pretschner}.} \bibinfo{year}{2022}\natexlab{}.
\newblock \showarticletitle{An empirical study on the effectiveness of static C code analyzers for vulnerability detection}. In \bibinfo{booktitle}{\emph{Proceedings of the 31st ACM SIGSOFT International Symposium on Software Testing and Analysis}}. \bibinfo{pages}{544--555}.
\newblock


\bibitem[Liu(2022)]%
        {diffmotivation}
\bibfield{author}{\bibinfo{person}{Dongmiao Liu}.} \bibinfo{year}{2022}\natexlab{}.
\newblock \bibinfo{booktitle}{\emph{SONARJAVA-73 add more lombok's used annotations for UnusedPrivateFieldCheck}}.
\newblock
\urldef\tempurl%
\url{https://github.com/SonarSource/sonar-java/pull/102#issuecomment-87545890}
\showURL{%
\tempurl}


\bibitem[Liu et~al\mbox{.}(2023)]%
        {issta2023sast}
\bibfield{author}{\bibinfo{person}{Han Liu}, \bibinfo{person}{Sen Chen}, \bibinfo{person}{Ruitao Feng}, \bibinfo{person}{Chengwei Liu}, \bibinfo{person}{Kaixuan Li}, \bibinfo{person}{Zhengzi Xu}, \bibinfo{person}{Liming Nie}, \bibinfo{person}{Yang Liu}, {and} \bibinfo{person}{Yixiang Chen}.} \bibinfo{year}{2023}\natexlab{}.
\newblock \showarticletitle{A Comprehensive Study on Quality Assurance Tools for Java}. In \bibinfo{booktitle}{\emph{Proceedings of the 32nd ACM SIGSOFT International Symposium on Software Testing and Analysis}} (Seattle, WA, USA) \emph{(\bibinfo{series}{ISSTA 2023})}. \bibinfo{publisher}{Association for Computing Machinery}, \bibinfo{address}{New York, NY, USA}, \bibinfo{pages}{285–297}.
\newblock
\showISBNx{9798400702211}
\urldef\tempurl%
\url{https://doi.org/10.1145/3597926.3598056}
\showDOI{\tempurl}


\bibitem[Liu et~al\mbox{.}(2022)]%
        {saner2022deepanna}
\bibfield{author}{\bibinfo{person}{Yi Liu}, \bibinfo{person}{Yadong Yan}, \bibinfo{person}{Chaofeng Sha}, \bibinfo{person}{Xin Peng}, \bibinfo{person}{Bihuan Chen}, {and} \bibinfo{person}{Chong Wang}.} \bibinfo{year}{2022}\natexlab{}.
\newblock \showarticletitle{DeepAnna: Deep Learning based Java Annotation Recommendation and Misuse Detection}. In \bibinfo{booktitle}{\emph{2022 IEEE International Conference on Software Analysis, Evolution and Reengineering (SANER)}}. IEEE, \bibinfo{pages}{685--696}.
\newblock


\bibitem[LynnBroe(2021)]%
        {pmd4487}
\bibfield{author}{\bibinfo{person}{LynnBroe}.} \bibinfo{year}{2021}\natexlab{}.
\newblock \bibinfo{booktitle}{\emph{UnnecessaryConstructor: false-positive with @Inject}}.
\newblock
\urldef\tempurl%
\url{https://github.com/pmd/pmd/issues/4487}
\showURL{%
\tempurl}


\bibitem[Mansur et~al\mbox{.}(2021)]%
        {fse2021datalog}
\bibfield{author}{\bibinfo{person}{Muhammad~Numair Mansur}, \bibinfo{person}{Maria Christakis}, {and} \bibinfo{person}{Valentin W{\"u}stholz}.} \bibinfo{year}{2021}\natexlab{}.
\newblock \showarticletitle{Metamorphic testing of Datalog engines}. In \bibinfo{booktitle}{\emph{Proceedings of the 29th ACM Joint Meeting on European Software Engineering Conference and Symposium on the Foundations of Software Engineering}}. \bibinfo{pages}{639--650}.
\newblock


\bibitem[marcelmore(2021)]%
        {pmd2454}
\bibfield{author}{\bibinfo{person}{marcelmore}.} \bibinfo{year}{2021}\natexlab{}.
\newblock \bibinfo{booktitle}{\emph{UnusedPrivateMethod violation for disabled class in 6.23}}.
\newblock
\urldef\tempurl%
\url{https://github.com/pmd/pmd/issues/2454}
\showURL{%
\tempurl}


\bibitem[martin(2020)]%
        {pmd2876}
\bibfield{author}{\bibinfo{person}{martin}.} \bibinfo{year}{2020}\natexlab{}.
\newblock \bibinfo{booktitle}{\emph{UnusedPrivateField cannot override ignored annotations property}}.
\newblock
\urldef\tempurl%
\url{https://github.com/pmd/pmd/issues/2876}
\showURL{%
\tempurl}


\bibitem[mjustin(2020)]%
        {cstyle7522}
\bibfield{author}{\bibinfo{person}{mjustin}.} \bibinfo{year}{2020}\natexlab{}.
\newblock \bibinfo{booktitle}{\emph{Exception when using SuppressWarningsHolder with @SuppressWarnings as an annotation property}}.
\newblock
\urldef\tempurl%
\url{https://github.com/checkstyle/checkstyle/issues/7522}
\showURL{%
\tempurl}


\bibitem[Mordahl et~al\mbox{.}(2023)]%
        {icse2023soottesting}
\bibfield{author}{\bibinfo{person}{Austin Mordahl}, \bibinfo{person}{Zenong Zhang}, \bibinfo{person}{Dakota Soles}, {and} \bibinfo{person}{Shiyi Wei}.} \bibinfo{year}{2023}\natexlab{}.
\newblock \showarticletitle{ECSTATIC: An Extensible Framework for Testing and Debugging Configurable Static Analysis}. In \bibinfo{booktitle}{\emph{2023 IEEE/ACM 45th International Conference on Software Engineering (ICSE)}}. \bibinfo{pages}{550--562}.
\newblock
\urldef\tempurl%
\url{https://doi.org/10.1109/ICSE48619.2023.00056}
\showDOI{\tempurl}


\bibitem[msridhar(2017)]%
        {infer559}
\bibfield{author}{\bibinfo{person}{msridhar}.} \bibinfo{year}{2017}\natexlab{}.
\newblock \bibinfo{booktitle}{\emph{Eradicate not reading annotations from class file}}.
\newblock
\urldef\tempurl%
\url{https://github.com/facebook/infer/issues/559}
\showURL{%
\tempurl}


\bibitem[Nedzelska(2021)]%
        {sonar3804}
\bibfield{author}{\bibinfo{person}{Marharyta Nedzelska}.} \bibinfo{year}{2021}\natexlab{}.
\newblock \bibinfo{booktitle}{\emph{FP in S3077 when volatile is used with @Immutable and @ThreadSafe annotations}}.
\newblock
\urldef\tempurl%
\url{https://sonarsource.atlassian.net/browse/SONARJAVA-3804}
\showURL{%
\tempurl}


\bibitem[nrmancuso(2020)]%
        {cstyle8734}
\bibfield{author}{\bibinfo{person}{nrmancuso}.} \bibinfo{year}{2020}\natexlab{}.
\newblock \bibinfo{booktitle}{\emph{Compact Constructor AST is missing annotations}}.
\newblock
\urldef\tempurl%
\url{https://github.com/checkstyle/checkstyle/issues/8734}
\showURL{%
\tempurl}


\bibitem[Nuryyev et~al\mbox{.}(2022a)]%
        {icsme2022mining}
\bibfield{author}{\bibinfo{person}{Batyr Nuryyev}, \bibinfo{person}{Ajay~Kumar Jha}, \bibinfo{person}{Sarah Nadi}, \bibinfo{person}{Yee-Kang Chang}, \bibinfo{person}{Emily Jiang}, {and} \bibinfo{person}{Vijay Sundaresan}.} \bibinfo{year}{2022}\natexlab{a}.
\newblock \showarticletitle{Mining Annotation Usage Rules: A Case Study with MicroProfile}. In \bibinfo{booktitle}{\emph{2022 38th International Conference on Software Maintenance and Evolution, IEEE}}.
\newblock


\bibitem[Nuryyev et~al\mbox{.}(2022b)]%
        {nuryyev2022mining}
\bibfield{author}{\bibinfo{person}{Batyr Nuryyev}, \bibinfo{person}{Ajay~Kumar Jha}, \bibinfo{person}{Sarah Nadi}, \bibinfo{person}{Yee-Kang Chang}, \bibinfo{person}{Emily Jiang}, {and} \bibinfo{person}{Vijay Sundaresan}.} \bibinfo{year}{2022}\natexlab{b}.
\newblock \showarticletitle{Mining Annotation Usage Rules: A Case Study with MicroProfile}. In \bibinfo{booktitle}{\emph{2022 IEEE International Conference on Software Maintenance and Evolution (ICSME)}}. IEEE, \bibinfo{pages}{553--562}.
\newblock


\bibitem[Olivera(2017)]%
        {googlejsr305}
\bibfield{author}{\bibinfo{person}{Fernando~Rodriguez Olivera}.} \bibinfo{year}{2017}\natexlab{}.
\newblock \bibinfo{booktitle}{\emph{FindBugs JSR305}}.
\newblock
\urldef\tempurl%
\url{https://mvnrepository.com/artifact/com.google.code.findbugs/jsr305}
\showURL{%
\tempurl}


\bibitem[Olivera(2023)]%
        {mvncentral}
\bibfield{author}{\bibinfo{person}{Fernando~Rodriguez Olivera}.} \bibinfo{year}{2023}\natexlab{}.
\newblock \bibinfo{booktitle}{\emph{MvnRepository}}.
\newblock
\urldef\tempurl%
\url{https://mvnrepository.com/}
\showURL{%
\tempurl}


\bibitem[oowekyala(2022)]%
        {pmdworkflow}
\bibfield{author}{\bibinfo{person}{oowekyala}.} \bibinfo{year}{2022}\natexlab{}.
\newblock \bibinfo{booktitle}{\emph{How PMD Works}}.
\newblock
\urldef\tempurl%
\url{https://docs.pmd-code.org/pmd-doc-6.53.0/pmd_devdocs_how_pmd_works.html}
\showURL{%
\tempurl}


\bibitem[Parnin et~al\mbox{.}(2013)]%
        {parnin2013adoption}
\bibfield{author}{\bibinfo{person}{Chris Parnin}, \bibinfo{person}{Christian Bird}, {and} \bibinfo{person}{Emerson Murphy-Hill}.} \bibinfo{year}{2013}\natexlab{}.
\newblock \showarticletitle{Adoption and use of Java generics}.
\newblock \bibinfo{journal}{\emph{Empirical Software Engineering}} \bibinfo{volume}{18}, \bibinfo{number}{6} (\bibinfo{year}{2013}), \bibinfo{pages}{1047--1089}.
\newblock


\bibitem[Peru(2015)]%
        {sonar1420}
\bibfield{author}{\bibinfo{person}{Nicolas Peru}.} \bibinfo{year}{2015}\natexlab{}.
\newblock \bibinfo{booktitle}{\emph{Annotation on array type should be properly handled}}.
\newblock
\urldef\tempurl%
\url{https://sonarsource.atlassian.net/browse/SONARJAVA-1420}
\showURL{%
\tempurl}


\bibitem[Pinheiro et~al\mbox{.}(2020)]%
        {pinheiro2020mutating}
\bibfield{author}{\bibinfo{person}{Pedro Pinheiro}, \bibinfo{person}{Jos{\'e}~Carlos Viana}, \bibinfo{person}{M{\'a}rcio Ribeiro}, \bibinfo{person}{Leo Fernandes}, \bibinfo{person}{Fabiano Ferrari}, \bibinfo{person}{Rohit Gheyi}, {and} \bibinfo{person}{Baldoino Fonseca}.} \bibinfo{year}{2020}\natexlab{}.
\newblock \showarticletitle{Mutating code annotations: An empirical evaluation on Java and C\# programs}.
\newblock \bibinfo{journal}{\emph{Science of Computer Programming}}  \bibinfo{volume}{191} (\bibinfo{year}{2020}), \bibinfo{pages}{102418}.
\newblock


\bibitem[PMD(2022)]%
        {pmdlink}
\bibfield{author}{\bibinfo{person}{PMD}.} \bibinfo{year}{2022}\natexlab{}.
\newblock \bibinfo{booktitle}{\emph{PMD An extensible cross-language static code analyzer.}}
\newblock
\urldef\tempurl%
\url{https://pmd.github.io/}
\showURL{%
\tempurl}


\bibitem[Pugh(2022)]%
        {jsr305dormant}
\bibfield{author}{\bibinfo{person}{William Pugh}.} \bibinfo{year}{2022}\natexlab{}.
\newblock \bibinfo{booktitle}{\emph{JSR 305: Annotations for Software Defect Detection}}.
\newblock
\urldef\tempurl%
\url{https://jcp.org/en/jsr/detail?id=305}
\showURL{%
\tempurl}


\bibitem[rnveach(2016)]%
        {cstyle3238}
\bibfield{author}{\bibinfo{person}{rnveach}.} \bibinfo{year}{2016}\natexlab{}.
\newblock \bibinfo{booktitle}{\emph{Java 8 Grammar: annotations on varargs parameters}}.
\newblock
\urldef\tempurl%
\url{https://github.com/checkstyle/checkstyle/issues/3238}
\showURL{%
\tempurl}


\bibitem[rnveach(2021)]%
        {cstyle9941}
\bibfield{author}{\bibinfo{person}{rnveach}.} \bibinfo{year}{2021}\natexlab{}.
\newblock \bibinfo{booktitle}{\emph{AtclauseOrder: Falsely ignores method with annotation}}.
\newblock
\urldef\tempurl%
\url{https://github.com/checkstyle/checkstyle/issues/9941}
\showURL{%
\tempurl}


\bibitem[robtimus(2021)]%
        {cstyle10945}
\bibfield{author}{\bibinfo{person}{robtimus}.} \bibinfo{year}{2021}\natexlab{}.
\newblock \bibinfo{booktitle}{\emph{OperatorWrap with token ASSIGN too strict for annotations}}.
\newblock
\urldef\tempurl%
\url{https://github.com/checkstyle/checkstyle/issues/10945}
\showURL{%
\tempurl}


\bibitem[Rocha and Valente(2011)]%
        {rocha2011annotations}
\bibfield{author}{\bibinfo{person}{Henrique Rocha} {and} \bibinfo{person}{Marco~Tulio Valente}.} \bibinfo{year}{2011}\natexlab{}.
\newblock \showarticletitle{How Annotations are Used in Java: An Empirical Study.}. In \bibinfo{booktitle}{\emph{SEKE}}. \bibinfo{pages}{426--431}.
\newblock


\bibitem[Shen et~al\mbox{.}(2021)]%
        {shen2021dlbugstudy}
\bibfield{author}{\bibinfo{person}{Qingchao Shen}, \bibinfo{person}{Haoyang Ma}, \bibinfo{person}{Junjie Chen}, \bibinfo{person}{Yongqiang Tian}, \bibinfo{person}{Shing-Chi Cheung}, {and} \bibinfo{person}{Xiang Chen}.} \bibinfo{year}{2021}\natexlab{}.
\newblock \showarticletitle{A comprehensive study of deep learning compiler bugs}. In \bibinfo{booktitle}{\emph{Proceedings of the 29th ACM Joint meeting on european software engineering conference and symposium on the foundations of software engineering}}. \bibinfo{pages}{968--980}.
\newblock


\bibitem[SonarQube(2022)]%
        {sonarqubelink}
\bibfield{author}{\bibinfo{person}{SonarQube}.} \bibinfo{year}{2022}\natexlab{}.
\newblock \bibinfo{booktitle}{\emph{SonarQube Code Quality and Code Security}}.
\newblock
\urldef\tempurl%
\url{https://www.sonarqube.org/}
\showURL{%
\tempurl}


\bibitem[Soot(2023)]%
        {sootlink}
\bibfield{author}{\bibinfo{person}{Soot}.} \bibinfo{year}{2023}\natexlab{}.
\newblock \bibinfo{booktitle}{\emph{Soot - A framework for analyzing and transforming Java and Android applications}}.
\newblock
\urldef\tempurl%
\url{http://soot-oss.github.io/soot/}
\showURL{%
\tempurl}


\bibitem[SpotBugs(2022)]%
        {spotbugslink}
\bibfield{author}{\bibinfo{person}{SpotBugs}.} \bibinfo{year}{2022}\natexlab{}.
\newblock \bibinfo{booktitle}{\emph{SpotBugs: Find bugs in Java Programs}}.
\newblock
\urldef\tempurl%
\url{https://spotbugs.github.io/}
\showURL{%
\tempurl}


\bibitem[Sun et~al\mbox{.}(2016)]%
        {issta2016gccstudy}
\bibfield{author}{\bibinfo{person}{Chengnian Sun}, \bibinfo{person}{Vu Le}, \bibinfo{person}{Qirun Zhang}, {and} \bibinfo{person}{Zhendong Su}.} \bibinfo{year}{2016}\natexlab{}.
\newblock \showarticletitle{Toward understanding compiler bugs in GCC and LLVM}. In \bibinfo{booktitle}{\emph{Proceedings of the 25th international symposium on software testing and analysis}}. \bibinfo{pages}{294--305}.
\newblock


\bibitem[Sun et~al\mbox{.}(2017)]%
        {sun2017mlbug}
\bibfield{author}{\bibinfo{person}{Xiaobing Sun}, \bibinfo{person}{Tianchi Zhou}, \bibinfo{person}{Gengjie Li}, \bibinfo{person}{Jiajun Hu}, \bibinfo{person}{Hui Yang}, {and} \bibinfo{person}{Bin Li}.} \bibinfo{year}{2017}\natexlab{}.
\newblock \showarticletitle{An empirical study on real bugs for machine learning programs}. In \bibinfo{booktitle}{\emph{2017 24th Asia-Pacific Software Engineering Conference (APSEC)}}. IEEE, \bibinfo{pages}{348--357}.
\newblock


\bibitem[Taneja et~al\mbox{.}(2020)]%
        {taneja2020testing}
\bibfield{author}{\bibinfo{person}{Jubi Taneja}, \bibinfo{person}{Zhengyang Liu}, {and} \bibinfo{person}{John Regehr}.} \bibinfo{year}{2020}\natexlab{}.
\newblock \showarticletitle{Testing static analyses for precision and soundness}. In \bibinfo{booktitle}{\emph{Proceedings of the 18th ACM/IEEE International Symposium on Code Generation and Optimization}}. \bibinfo{pages}{81--93}.
\newblock


\bibitem[Tang et~al\mbox{.}(2010)]%
        {jtres2010static}
\bibfield{author}{\bibinfo{person}{Daniel Tang}, \bibinfo{person}{Ales Plsek}, {and} \bibinfo{person}{Jan Vitek}.} \bibinfo{year}{2010}\natexlab{}.
\newblock \showarticletitle{Static checking of safety critical Java annotations}. In \bibinfo{booktitle}{\emph{Proceedings of the 8th International Workshop on Java Technologies for Real-Time and Embedded Systems}}. \bibinfo{pages}{148--154}.
\newblock


\bibitem[Thung et~al\mbox{.}(2012)]%
        {issre2012mlsystem}
\bibfield{author}{\bibinfo{person}{Ferdian Thung}, \bibinfo{person}{Shaowei Wang}, \bibinfo{person}{David Lo}, {and} \bibinfo{person}{Lingxiao Jiang}.} \bibinfo{year}{2012}\natexlab{}.
\newblock \showarticletitle{An empirical study of bugs in machine learning systems}. In \bibinfo{booktitle}{\emph{2012 IEEE 23rd International Symposium on Software Reliability Engineering}}. IEEE, \bibinfo{pages}{271--280}.
\newblock


\bibitem[Tomassi and Rubio-Gonz{\'a}lez(2021)]%
        {ase2021npe}
\bibfield{author}{\bibinfo{person}{David~A Tomassi} {and} \bibinfo{person}{Cindy Rubio-Gonz{\'a}lez}.} \bibinfo{year}{2021}\natexlab{}.
\newblock \showarticletitle{On the real-world effectiveness of static bug detectors at finding null pointer exceptions}. In \bibinfo{booktitle}{\emph{2021 36th IEEE/ACM International Conference on Automated Software Engineering (ASE)}}. IEEE, \bibinfo{pages}{292--303}.
\newblock


\bibitem[Vieira et~al\mbox{.}(2010)]%
        {cohen}
\bibfield{author}{\bibinfo{person}{Susana~M Vieira}, \bibinfo{person}{Uzay Kaymak}, {and} \bibinfo{person}{Jo{\~a}o~MC Sousa}.} \bibinfo{year}{2010}\natexlab{}.
\newblock \showarticletitle{Cohen's kappa coefficient as a performance measure for feature selection}. In \bibinfo{booktitle}{\emph{International conference on fuzzy systems}}. IEEE, \bibinfo{pages}{1--8}.
\newblock


\bibitem[Villard(2020)]%
        {inferworkflow}
\bibfield{author}{\bibinfo{person}{Jules Villard}.} \bibinfo{year}{2020}\natexlab{}.
\newblock \bibinfo{booktitle}{\emph{Infer workflow}}.
\newblock
\urldef\tempurl%
\url{https://fbinfer.com/docs/infer-workflow}
\showURL{%
\tempurl}


\bibitem[vovkss(2018)]%
        {pmd1369}
\bibfield{author}{\bibinfo{person}{vovkss}.} \bibinfo{year}{2018}\natexlab{}.
\newblock \bibinfo{booktitle}{\emph{[java] Processing error (ClassCastException) if a TYPE\_USE annotation is used on a base class in the "extends" clause}}.
\newblock
\urldef\tempurl%
\url{https://github.com/pmd/pmd/issues/1369}
\showURL{%
\tempurl}


\bibitem[Wan et~al\mbox{.}(2017)]%
        {msr2017blockchain}
\bibfield{author}{\bibinfo{person}{Zhiyuan Wan}, \bibinfo{person}{David Lo}, \bibinfo{person}{Xin Xia}, {and} \bibinfo{person}{Liang Cai}.} \bibinfo{year}{2017}\natexlab{}.
\newblock \showarticletitle{Bug characteristics in blockchain systems: a large-scale empirical study}. In \bibinfo{booktitle}{\emph{2017 IEEE/ACM 14th International Conference on Mining Software Repositories (MSR)}}. IEEE, \bibinfo{pages}{413--424}.
\newblock


\bibitem[Wang et~al\mbox{.}(2022)]%
        {2022icpcfinder}
\bibfield{author}{\bibinfo{person}{Junjie Wang}, \bibinfo{person}{Yuchao Huang}, \bibinfo{person}{Song Wang}, {and} \bibinfo{person}{Qing Wang}.} \bibinfo{year}{2022}\natexlab{}.
\newblock \showarticletitle{Find Bugs in Static Bug Finders}. In \bibinfo{booktitle}{\emph{2022 IEEE/ACM 30th International Conference on Program Comprehension (ICPC)}}. \bibinfo{pages}{516--527}.
\newblock
\urldef\tempurl%
\url{https://doi.org/10.1145/3377811.3380380}
\showDOI{\tempurl}


\bibitem[Yu et~al\mbox{.}(2021)]%
        {tse2019annotation}
\bibfield{author}{\bibinfo{person}{Zhongxing Yu}, \bibinfo{person}{Chenggang Bai}, \bibinfo{person}{Lionel Seinturier}, {and} \bibinfo{person}{Martin Monperrus}.} \bibinfo{year}{2021}\natexlab{}.
\newblock \showarticletitle{Characterizing the Usage, Evolution and Impact of Java Annotations in Practice}.
\newblock \bibinfo{journal}{\emph{IEEE Transactions on Software Engineering}} \bibinfo{volume}{47}, \bibinfo{number}{5} (\bibinfo{year}{2021}), \bibinfo{pages}{969--986}.
\newblock
\urldef\tempurl%
\url{https://doi.org/10.1109/TSE.2019.2910516}
\showDOI{\tempurl}


\bibitem[Zhang et~al\mbox{.}(2023)]%
        {statfier}
\bibfield{author}{\bibinfo{person}{Huaien Zhang}, \bibinfo{person}{Yu Pei}, \bibinfo{person}{Junjie Chen}, {and} \bibinfo{person}{Shin~Hwei Tan}.} \bibinfo{year}{2023}\natexlab{}.
\newblock \showarticletitle{Statfier: Automated Testing of Static Analyzers via Semantic-preserving Program Transformations}. In \bibinfo{booktitle}{\emph{Proceedings of the 31st ACM Joint European Software Engineering Conference and Symposium on the Foundations of Software Engineering}}. Association for Computing Machinery (ACM).
\newblock


\bibitem[Zhang et~al\mbox{.}(2020)]%
        {zhang2020dljob}
\bibfield{author}{\bibinfo{person}{Ru Zhang}, \bibinfo{person}{Wencong Xiao}, \bibinfo{person}{Hongyu Zhang}, \bibinfo{person}{Yu Liu}, \bibinfo{person}{Haoxiang Lin}, {and} \bibinfo{person}{Mao Yang}.} \bibinfo{year}{2020}\natexlab{}.
\newblock \showarticletitle{An empirical study on program failures of deep learning jobs}. In \bibinfo{booktitle}{\emph{Proceedings of the ACM/IEEE 42nd International Conference on Software Engineering}}. \bibinfo{pages}{1159--1170}.
\newblock


\bibitem[Zhang et~al\mbox{.}(2018)]%
        {zhang2018tfbug}
\bibfield{author}{\bibinfo{person}{Yuhao Zhang}, \bibinfo{person}{Yifan Chen}, \bibinfo{person}{Shing-Chi Cheung}, \bibinfo{person}{Yingfei Xiong}, {and} \bibinfo{person}{Lu Zhang}.} \bibinfo{year}{2018}\natexlab{}.
\newblock \showarticletitle{An empirical study on TensorFlow program bugs}. In \bibinfo{booktitle}{\emph{Proceedings of the 27th ACM SIGSOFT International Symposium on Software Testing and Analysis}}. \bibinfo{pages}{129--140}.
\newblock


\end{thebibliography}
